# Progress in epitaxial thin-film Na$_3$Bi as a topological electronic material

I. Di Bernardo,[1,2] J. Hellerstedt,[2] C. Liu,[1,2] G. Akhgar,[1,2] W. Wu,[3] S. A. Yang,[3] D. Culcer,[1,4] S.-K. Mo,[5] S. Adam,[6] M. T. Edmonds,[1,2,7*] M. S. Fuhrer[1,2,7*]

[1] *Australian Research Council Centre of Excellence in Future Low-Energy Electronics Technologies, Monash University, 3800 Clayton, Victoria, Australia*

[2] *School of Physics and Astronomy, Monash University, 3800 Clayton, Victoria, Australia*
[3] *Research Laboratory for Quantum Materials, Singapore University of Technology and Design, Singapore 487372, Singapore*
[4] *School of Physics, University of New South Wales, Sydney, New South Wales 2052 Australia*
[5] *Advanced Light Source, Lawrence Berkeley National Laboratory, Berkeley, California 94720, USA*
[6] *Yale-NUS College, 16 College Ave West, 138527 Singapore*
[7] *Monash Centre for Atomically Thin Materials, Monash University, 3800 Clayton Victoria, Australia*

**\* Corresponding author: mark.edmonds@monash.edu and michael.fuhrer@monash.edu**

**Abstract:**
Na$_3$Bi was the first experimentally verified topological Dirac semimetal (TDS), and is a 3D analogue of graphene hosting relativistic Dirac fermions. Its unconventional momentum-energy relationship is interesting from a fundamental perspective, yielding exciting physical properties such as chiral charge carriers, the chiral anomaly, and weak anti-localization. It also shows promise for realising topological electronic devices such as topological transistors. In this review, an overview of the substantial progress achieved in the last few years on Na$_3$Bi is presented, with a focus on technologically relevant large-area thin films synthesised via molecular beam epitaxy. Key theoretical aspects underpinning the unique electronic properties of Na$_3$Bi are introduced. Next, the growth process on different substrates is reviewed. Spectroscopic and microscopic features are illustrated, and an analysis of semi-classical and quantum transport phenomena in different doping regimes is provided. The emergent properties arising from confinement in two dimensions, including thickness-dependent and electric-field driven topological phase transitions, are addressed, with an outlook towards current challenges and expected future progress.

## 1. Introduction

Historically, materials have been divided into two classes according to their band structures: metals and insulators. In metals the chemical potential crosses the conduction band, giving non-zero conductance at any finite temperature, whilst in insulators it is located in a finite bandgap between the conduction and the valence band. With the discovery of the quantum Hall effect in two-dimensional (2D) systems subject to a perpendicular magnetic field,[1] a new class of *topological materials* has been identified, and since then the topological phases of matter have been subject of an ever-increasing number of studies. Insulators are further classified according to *topological invariants* which describe their band structure. Strikingly, it was found that even in the absence of a magnetic field or magnetization (i.e. in the presence of time-reversal symmetry) insulators are described by a topological invariant $\upsilon \in \mathbb{Z}_2$; that is, there are *two* types of time-reversal symmetric insulators. The invariant $\upsilon = 0$ corresponds to a conventional insulator, and $v = 1$ corresponds to a *topological insulator* where in the presence of a strong spin orbit coupling, the valence and conduction band invert.[2–6]

An important aspect of topological order is that the topological invariant describing a state cannot be changed as long as the bandgap remains open and the system symmetry is preserved. This leads to a *bulk-boundary correspondence*: At the interface between two states of different topological invariant (for example topological and conventional insulator) there will be a metallic gapless state. These edge states (in 2D) or surface states (in three dimensions; 3D) are generically Dirac-like in nature, with oppositely dispersing bands crossing at a Dirac point about which the dispersion is linear. Thus, the boundaries of 2D or 3D topological insulators with conventional insulators (or with vacuum) host gapless, metallic, Dirac edge or surface states. Any parameter (such as strain, chemical substitution, electric field) which tunes a material from conventional to topological insulator may have an intervening gapless state, making the material a *topological Dirac semimetal* (TDS).

In 3D and in case of linear dispersion in all directions, the Dirac semimetal is characterised by *Dirac points* (consisting of degenerate *Weyl points*) at which the system Hamiltonian takes the form $\hat{H}(\boldsymbol{k}) = v_{ij} k_i \sigma_j$, where $\vec{\sigma} = \{\sigma_x, \sigma_y, \sigma_z\}$ are the Pauli matrices. Imposing a non-vanishing condition on $\det[v_{ij}]$ ensures robustness against perturbations. This Hamiltonian describes two linear bands, degenerate at the Weyl point (or node). The robustness of the node is associated with the Chern number of a sphere encircling the valence band at the node, $sign(\det[v_{ij}]) = \pm 1$. Time-reversal symmetry (TRS) requires that for every Weyl node occurring at $\boldsymbol{k}_W$ in momentum space, another occurs at $-\boldsymbol{k}_W$ with the same Chern number. On the other hand, inversion symmetry (IS) requires the Weyl points at $\boldsymbol{k}_W$ and $-\boldsymbol{k}_W$ to have opposite Chern number. Therefore, in time-reversal and inversion invariant systems, isolated Weyl points cannot exist, instead, pairs of oppositely charged Weyl points have to merge to form fourfold degenerate linear *Dirac* points. In other words, Dirac fermions are degenerate Weyl fermions with opposite chirality. Weyl semimetals, with twofold Weyl points, can be obtained by breaking TRS or IS in a Dirac semimetal. TRS-breaking Weyl semimetals are uncommon: their existence was predicted only in few magnetically ordered materials – like pyrochlore iridates[7] ($Y_2Ir_2O_7$) and Co-based Heusler materials,[8,9] and verified experimentally only very recently.[10–12] The family of IS-breaking materials is much broader, with theoretical predictions confirmed experimentally, among others (like TaAs[13–15]), for the transition metal dichalcogenides $WTe_2$[16–18] and $MoTe_2$.[19,20]

Three dimensional Dirac semimetals were first proposed to arise from accidental degeneracies such as a topological phase transition from a conventional to a topological insulator. Accidental Dirac materials have been both predicted in theoretical calculations,[21–23] and verified experimentally.[24–30] In general, though, the accidental degeneracy requires fine-tuning of parameters, as the superposition of two Weyl nodes generically results in band hybridization and gap opening. This difficulty can be removed by protecting the degeneracy by a further space-group symmetry. The criteria to identify which space groups can host relativistic Dirac fermions, along with theoretical predictions on stable 3D Dirac semimetals are outlined in Ref. [31] and [32].

The presence of Dirac or Weyl points leads to unusual and novel electronic phenomena. The surface states of Weyl semimetals are open Fermi arcs, connecting the surface projections of bulk Weyl nodes with opposite chiralities. These Fermi arc states have been imaged via angle-resolved photoemission spectroscopy (ARPES).[7,13,17,20,33–35] and are the cause of various transport phenomena, such as unusual Shubnikov-de-Haas oscillations[33,36,37] or the unconventional quantum Hall effect.[35,38–40] As TDS are degenerate Weyl semimetals, Fermi arcs are expected (and have indeed been observed) for this class of materials,[33,41,42] even though their topological robustness has been questioned.[43,44] The presence of Dirac or Weyl points also offers a condensed matter realization of the chiral anomaly, i.e., the charge pumping between nodes of opposite chiralities in the presence of parallel electric and magnetic fields. [45–48]

In 2012 Young et al.[31] first predicted the existence of three-dimensional Dirac semimetals and speculated β-cristobalite $BiO_2$ to be a prototypical example, but it remains unrealised experimentally to date. Shortly after, Wang et al.[32] predicted, based on first principles calculations, that crystalline, stoichiometric compounds $A_3Bi$ (A= Na, K, Rb) were also three-dimensional Dirac semimetals, with bulk Dirac points protected by crystal symmetry, and that topological phase transitions could be easily driven in these materials. This was followed by the first experimental demonstration of a Dirac semimetal in $Na_3Bi$ via imaging of the Dirac points using ARPES.[49] Another Dirac semimetal $Cd_3As_2$ was also predicted theoretically[50] and observed experimentally.[51] Like $Na_3Bi$, $Cd_3As_2$ is also amenable to controlled growth via molecular beam epitaxy[52,53] and has exhibited numerous signatures of topological electronic structure. [33,48,54–56] A detailed description of $Cd_3As_2$ material properties can be found in Ref. [57].

Compared to $Cd_3As_2$, with up to 80 atoms per unit cell,[50] $Na_3Bi$ is structurally simple, with eight atoms per unit cell and only two symmetry protected Dirac nodes in proximity of the Γ point. These linearly dispersing bands are isolated over a wide range of energies around the Fermi level. This simplifies theoretical modelling and allows easy experimental exploration of the band structure in proximity to the nodes, with access to different transport regimes readily achieved via simple gating or doping. Several studies predicted and confirmed its TDS nature by probing different aspects of its electronic structure: evidence of the chiral anomaly was detected in $Na_3Bi$,[46,47] as well as perfect weak antilocalization – a clear signature of spin-momentum locking and Dirac fermions.[58,59] Furthermore, $Na_3Bi$ can be grown epitaxially with little interaction on a variety of conductive substrates such as silicon, [60–63] and graphene,[63–65] or insulating substrates such as sapphire,[59,60,66–69] so that $Na_3Bi$-based devices could be employed in a broad range of applications. Most importantly, ultrathin $Na_3Bi$ was the first material to demonstrate reversible electric-field driven switching of its topological phase.[70] For these reasons, $Na_3Bi$ stands as the most advanced material system towards the realisation of practical topological electronics.

In this progress report we systematically present the recent theoretical and experimental developments concerning MBE-grown thin and ultrathin films of $Na_3Bi$. In Section 2 we present the numerical results and the computed band structure: we introduce the lattice parameters and a simple model for the derivation of the band structure, which is strongly affected by spin-orbit coupling effects. We subsequently compare the results obtained for the dispersion with different theoretical approaches. Section 3 discusses the experimentally determined electronic properties of thin film samples: after an overview of the growth methods on different substrates, we describe the atomic structure in terms of scanning tunnelling microscopy measurements, and the band structure as obtained by photoemission and scanning tunnelling spectroscopy. A detailed analysis of the transport properties of the Na3Bi, both in the semiclassical and in the quantum transport regime is then provided. Section 4 introduces the recent discoveries about the effect of 2D confinement on the bandgap size and the topological character of the material, followed by a description of the electric-field-driven topological phase transition and the most recent results for edge state transport. Prospective applications of Na3Bi and current challenges for the large-scale production of devices (ranging from scalability to air stability) are addressed in Section 5.

## 2. Material fundamentals: numerical results and simulations

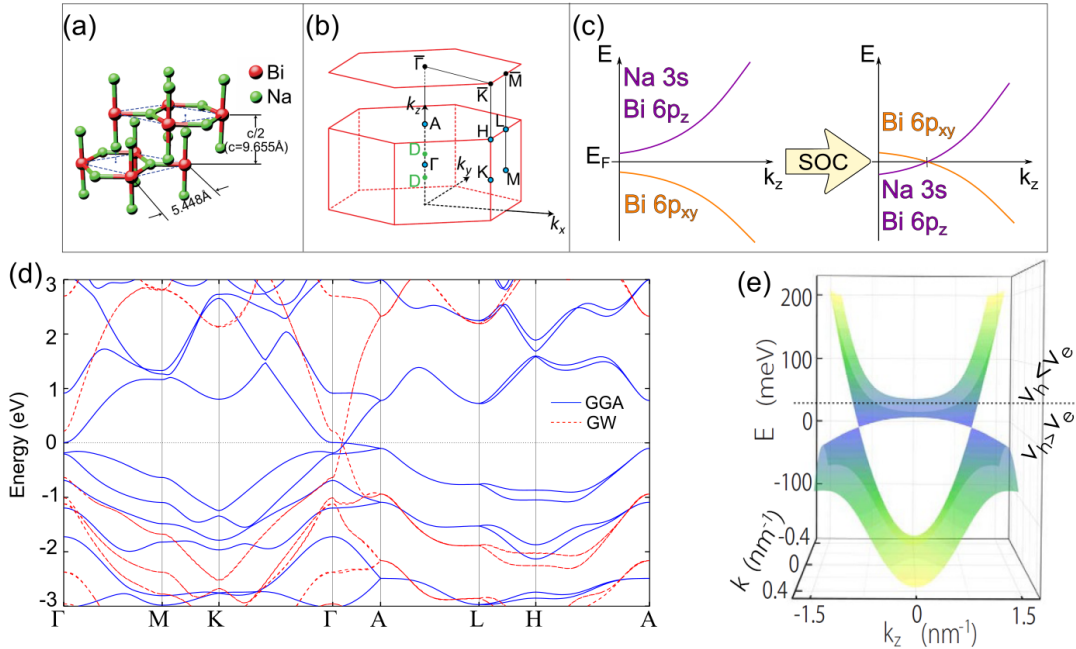

*Figure 1:* First-principle calculations of Na3Bi band structure. (a) Crystal structure and (b) 3D Brillouin zone projected on the (001) surface. (c) Sketch of the spin-orbit coupling induced band inversion mechanism.(d) Comparison of the calculated band structures for Na3Bi with GGA (blue solid lines), and GW (red dashed lines) approaches along the main symmetry directions (e) Energy spectrum plotted as a function of $k_z$ and $k_\parallel$, obtained from the model in Eq.(1). Panels (a), (b) Reprinted from Zhang et al., Appl. Phys. Lett. 105, 031901 (2014), with the permission of AIP Publishing. Panel (d) adapted with permission from Di Bernardo, I., et al., Phys. Rev. B 102, 045124 (2020). Copyright (2020) by the American Physical Society. Panel (e) adapted with permission from X. Xiao et al., Sci. Rep 5:7898.

Na3Bi belongs to the alkali pnictides family $A_3B$ (A= alkali, B= As, Sb, Bi) and, like $K_3Bi$ and $Rb_3Bi$, crystallizes in a hexagonal $P6_3/mmc$ structure[62,64] [see Fig. 1(a)], with two non-equivalent sites for sodium: Na(1) and Bi atoms form honeycomb lattice layers spaced by Na(2) atoms connecting two adjacent Bi sites.[32] The calculated in-plane lattice constant is $a = 5.45$Å, while the separation between consecutive Na-Bi layers is $c/2 = 4.8$Å ($c/2$ corresponds to a monolayer, while the full unit cell is constituted by a bilayer).[42,62–67,71] The corresponding three-dimensional Brillouin zone (BZ) is shown in Figure 1(b), as well as its 2D projection along the (001) plane, highlighting the high symmetry points.

The valence and conduction bands emerging from this atomic structure predominantly exhibit Bi-*6p* and Na-*3s* orbital character [about 65% from Na(1) and 35% from Na(2)],[32] with the top of the valence band dominated by Bi-*6p$_{x,y}$* states and the conduction band by Na-*3s* states. At the Γ− point the Na-*3s* band is calculated by GGA simulations to be 0.7 eV lower than the Bi-*6p$_{x,y}$* band because of spin-orbit contributions:[32] this band inversion, depicted in Figure 1(c), results in a semimetallic, non-gapped band structure, with band crossings (nodes) located exactly at the Fermi level in charge-neutral Na$_3$Bi. As the crystal structure is both time reversal (TR) and inversion (I) symmetric (as well as $C_3$ symmetric along the (001) axis), these nodes are classified as Dirac points (consisting of two Weyl points of opposite charge),[72,73] each characterized by four-fold degeneracy. Therefore, around the Fermi points the effective Hamiltonian can be linearized to obtain a Dirac cone-like linear dispersion.

A simple, low-energy effective model for Dirac semimetals was constructed following the first-principles calculations for A$_3$Bi (A = Na, K, Rb).[31,32,71] The states closest to the Fermi level at the Γ point consist of a four orbital basis listed from lowest to highest energy: $\left|S_{\frac{1}{2}}^+, \frac{1}{2}\right\rangle, \left|P_{\frac{3}{2}}^-, \frac{3}{2}\right\rangle, \left|S_{\frac{1}{2}}^+, -\frac{1}{2}\right\rangle, \left|P_{\frac{3}{2}}^-, -\frac{3}{2}\right\rangle$; the subscript indicates the total angular momentum |J| and the superscript indicates the parity of the orbital ($\left|S^\pm\right\rangle = \frac{1}{\sqrt{2}}(|Na, s\rangle \pm |Na', s\rangle)$, where Na'=I(Na) denotes equivalence under inversion symmetry). Around the Γ point up to second order in *k*, the effective $4x4$ $k \cdot p$ Hamiltonian can be written as:

$$H(\mathbf{k}) = \varepsilon_0(\mathbf{k}) + \begin{bmatrix} M(\mathbf{k}) & Ak_+ & 0 & 0 \\ Ak_- & -M(\mathbf{k}) & 0 & 0 \\ 0 & 0 & M(\mathbf{k}) & -Ak_- \\ 0 & 0 & -Ak_+ & -M(\mathbf{k}) \end{bmatrix} \quad (1)$$

Here, $\varepsilon_0(\mathbf{k}) = C_0 + C_1 k_z^2 + C_2(k_x^2 + k_y^2)$, $k_\pm = k_x \pm ik_y$, and $M(\mathbf{k}) = CM_0 - M_1 k_z^2 - M_2(k_x^2 + k_y^2)$. The parameters $A, C_i, M_i$ ($M_i < 0$ to reproduce the band inversion) are material specific and can be obtained by fitting *ab initio* calculations as in Ref. [74]. The block-Hamiltonian in Eq. (1) can be decoupled into two ($2x2$) matrices, one for each of the two overlapping Weyl points degenerate in energy with opposite chiralities.

The eigenvalues of this system give an energy spectrum of the form $E(\mathbf{k}) = \varepsilon_0(\mathbf{k}) \pm \sqrt{M(\mathbf{k})^2 + A^2(k_x^2 + k_y^2)}$, where two gapless solutions are obtained by constraining the term under the square-root to vanish. These Dirac points are located on the $k_z$ axis (Γ − A direction) symmetric with respect to Γ, at $\pm \mathbf{k}_D = (0, 0, \pm k_D)$ with $k_D = \sqrt{M_0/M_1} = 0.903\ nm^{-1}$. By expanding $E(\mathbf{k})$ around $\pm \mathbf{k}_D$ it is straightforward to verify that the dispersion is linear around the primary rotation axis and in all three directions. However, these linear dispersions are anisotropic– the Fermi velocity along $k_z$ is much lower than along $k_x - k_y$. The spectrum can be considered to be Dirac-like over the range $-180\ meV$ to $+30\ meV$ (estimated via HSE06 method), while the range over which the band spectrum only consists of the overlapping bands is $[-0.1, 1.2]\ eV$ according to HSE06 calculations[61] (however recent GW calculations and experiments suggest this may be significantly larger; see below).

Given the suppressed density of states at the Fermi level, the exchange and correlation (XC) interactions are expected to play a major role in determining its band structure. To shed light on this issue, a recent work by Di Bernardo et al.[61] reported first-principles theoretical calculations of the band structure using an increasing hierarchy of approximations of energy corrections due to interactions: generalized gradient approximation (GGA) with Perdew-Burke-Ernzerhof (PBE) potential,[75] SCAN meta-GGA,[76] the Heyd–Scuseria–Ernzerhof hybrid functional (HSE06),[77,78] as well as the GW approximation.[79] The GGA and GW results are reported in Figure 1(d). Density functional theory-based methods were found to yield qualitatively consistent band structures, with slight differences in the low energy region and larger Fermi velocities (especially for the electron band) with an increasingly sophisticated description of the XC potential. The GW estimate of the Fermi velocity was observed to be even larger ($v_z = 1.3 \cdot 10^6\ ms^{-1}$), highlighting the importance of considering exchange-correlation effects for this system.

From these calculations it is evident that Dirac cones form in the bulk of Na$_3$Bi along the $\Gamma - A$ direction [Figure 1(d)]. For clarity, in Figure 1(e) shows a three-dimensional model of the bulk band structure as a function of $k_z$ and $k_\parallel = \sqrt{k_x^2 + k_y^2}$, as obtained by Xiao et al.[61] The linear dispersion at the two Dirac nodes is clearly visible, as well as the merging of the two conduction and valence band cones away from the apex of the Dirac cone $E_D$. This merging is associated with a topological change in the Fermi surface, which evolves from two unconnected Dirac contours (one loop per Dirac point) to a single (diamond-shaped) loop.[80] This change from a *genus*-2 to a *genus*-1 surface is associated with a Lifshitz transition,[81,82] and connected to the appearance of a saddle point in the band structure.[80] In light of this phenomenon, we note that the band structure can be roughly divided into two sections, exhibiting different ratios between the slopes of the valence and conduction bands, or the ratio of the velocity of the charge carriers: in close proximity of the Dirac nodes and below the Lifshitz transition point ($|E - E_D| \leq 40\ meV$), the conduction band is flatter than the valence band, so in this regime holes are "faster" than electrons; further from the Dirac point the roles reverse with the conduction band becoming much steeper (about 7 times) than the valence band.

According to GW calculations, only two bands can be observed in the Na$_3$Bi band spectrum over the range $[-0.9, 2.1]\ eV$, while the Dirac-like approximation is valid for the conduction and valence bands (below/above Fermi) in the energy range between the Lifshitz transition energies of $-630\ meV$ and $220\ meV$.[61] These values, slightly different than the ones obtained via HSE, are considerably larger than those reported for Cd$_3$As$_2$, where the energy window in which only the Dirac bands can be observed is about $[-0.25, 0.25]\ eV$ (obtained via GGA calculations),[83] and the Lifshitz transitions energies are at $-30\ meV$ and $+10\ meV$ as extracted from PBE calculations.[84] In other words, Na$_3$Bi behaves like a Dirac semimetal over a much larger window around the Fermi level than Cd$_3$As$_2$.

# 3. Experimental results in thin-film Na$_3$Bi

In this section we focus on the experimental characterization of thin films of Na$_3$Bi grown via molecular beam epitaxy. After discussing the main issues associated with the growth process, we highlight the main findings in terms of topography and electronic structure. Finally, the electronic transport properties in high- and low carrier density regimes are reviewed.

## 3.1. Growth processes and characterization

Molecular beam epitaxy (MBE) in ultra-high vacuum has proven to be the best growth technique for Na$_3$Bi thin films due to the extreme volatility of Na, as it can be supplied in excess while maintaining perfect stoichiometry as first reported by Zhang et al.[63] At the typical growth temperatures (between 250°C and 390°C at a pressure in the range of $10^{-9}$ mbar),[60] Na:Bi ratios as large as 20:1 are necessary. In these conditions the bismuth rate is the main limiting factor for growth, as the growth temperature is above the sublimation temperature of Na meaning it does not stick to the surface.[64] The intrinsic volatility of sodium often results in surfaces rich with Na vacancies. In this regard, a post-growth anneal in pure Na overflux has been demonstrated to be effective in "curing" the surface by mitigating the presence of vacancies.[65]

Thin films of Na$_3$Bi were first explored in bilayer graphene on SiC[63–65] and Si(111) 7x7[60–63] as they are ultraflat, possess similar three-fold rotational symmetry, and had been used in the past to grow ultra-thin films of other topological insulators.[85–89] Zhang et al.[63] reported growth on both graphene and Si(111), where Na$_3$Bi was reported to grow with a 30° in-plane rotation with respect to the substrate (the Γ-K direction on the substrate corresponds to the Γ-M direction on Na$_3$Bi and vice versa), as observed subsequently in several other studies via reflection high-energy electron diffraction (RHEED).[59,62–65,67] Hellerstedt et al.[66] demonstrated that Na$_3$Bi can also be successfully grown on insulating substrates like Al$_2$O$_3$[0001] with a two-step process, involving the growth of a seed layer at 120°C followed by additional growth at higher temperatures.[66] While growth on this substrate exhibits increased levels of both point defects and morphological disorder, relatively large (>100nm) single-domain areas can still be achieved.[59,60,66–69] Film quality is sensitively dependent on the Al$_2$O$_3$ substrate temperature, as growth temperatures larger than 380°C result in film dewetting.[66]

On all reported substrates, MBE growth of Na$_3$Bi proceeds according to the Volmer-Weber model,[90] with the formation of multi-layer islands on the substrate before the completion of the first full monolayer. This means that the interaction between the evaporated species is stronger than their interaction with the substrate, a conclusion supported both by RHEED and STM measurements. During the growth of subsequent layers the intensity of the RHEED central spot was not observed to oscillate, consistent with Volmer-Weber (rather than layer-by-layer Frank-van der Merwe) growth.[62] Several works showing large-area STM scans of Na$_3$Bi, like the one in Figure 2(a), often reported the formation of atomically flat terraces with lateral extent of hundreds of nm,[59,62,64,67,91,92] sometimes of very different thicknesses.[65] The appearance of screw lines and hexagonal domains has been observed on all substrates,[60,64] while from atomic resolution images [inset of Figure 2(a)] a lattice constant of 5.45 Å can be measured.[60,61,66]

Recently, Pinchuck et al.[67] proposed alternating-layer MBE as a way to improve the overall smoothness of the films: in this work, growth on Al$_2$O$_3$[0001] was carried out at room temperature by alternating a flux-matched NaBi monolayer with a pure Na bilayer for the seeding layer. With this method, a roughness below 1 nm was achieved as measured by atomic force microscopy. Subsequent layers of Na$_3$Bi grown on these seeding layers are also reported to be more homogeneous and with a larger crystal size compared with those grown by simple Na-overflux MBE.

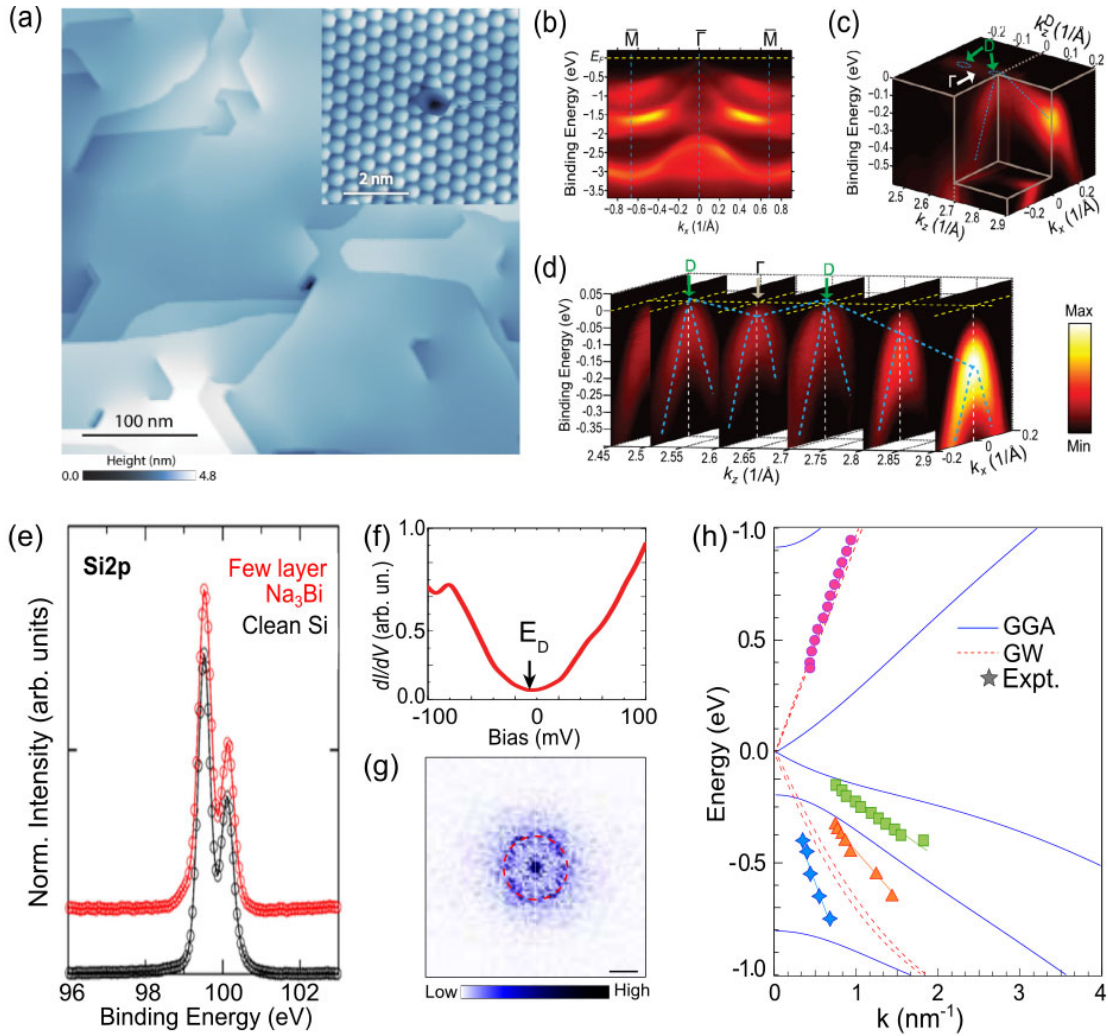

*Figure 2:* Microscopic and spectroscopic characterization of Na$_3$Bi films. (a) Large-area (400 nm × 380 nm) topographic STM image (bias voltage $V = -3V$ and tunnel current $I = 50$ pA) of Na$_3$Bi grown on Si(111). Inset: Atomic-resolution STM image taken on a separate Na$_3$Bi film showing a single Na vacancy at the surface.[60] (b)-(d): ARPES spectra of a thin film of

*Na₃Bi grown on bilayer graphene: (b) Low energy band structure along the $\bar{\Gamma} - \bar{M}$ direction (high symmetry points highlighted with blue dashed lines). (c) 3D intensity plot of the 4D ARPES spectra projected in the $k_x - k_y - E_B$ space at $k_y = 0$. Blue dashed lines serve as a guide to the eye to follow the linear dispersion of the top band around the Dirac point. The Γ and Dirac (D) points are highlighted by white and green arrows, respectively. (d) Stack of ARPES spectra in the $k_x - E_B$ plane at $k_y = 0$ along the $k_z$ direction. Yellow dashed lines indicate the Fermi energy ($E_F$), blue dashed lines follow the linear dispersion of the band top. (e) Si 2p core level of the clean Si substrate (black curve) and with few-layer Na₃Bi grown on top (red curve); $hv = 850\ eV$. (f) dI/dV spectrum of Na₃Bi on Al₂O₃[0001], where the position of the Dirac pint is indicated by a black arrow. (g) 2D FFT of a 60 nm x60 nm dI/dV image of Na₃Bi, where the scale bar is 2.1 nm⁻¹. (h) Calculated band structure for Na₃Bi along the $\bar{\Gamma} - \bar{K}$ direction in the (001) plane containing the Dirac point. Markers are the measured data from QPI, linear fittings to evaluate the Fermi velocities are superimposed on the data. Panel (a) adapted from Edmonds, M.T, et al., Sci. Adv. 2017, 3, eaao6661. Panels (b)-(d) reprinted from Zhang et al. Appl. Phys. Lett. 105, 031901 (2014)., with the permission of AIP Publishing. Panel (e) adapted from Collins, J., et al., Nature 2018, 564, 390. Panel (f) adapted with permission from Hellerstedt, J. et al., Nano Lett. 16(5), 3210–3214 (2016). Copyright (2016) American Chemical Society. Panels (g) and (h) adapted with permission from Di Bernardo, I., et al., Phys. Rev. B 102, 045124 (2020). Copyright (2020) by the American Physical Society.*

## 3.2. Electronic structure

Confirmation that Na₃Bi thin films less than 20 nm retain the electronic properties of a 3D topological Dirac semimetal was reported by Mo and co-workers using angle-resolved photoelectron spectroscopy (ARPES) for Na₃Bi grown epitaxially on bilayer graphene (BLG) on SiC.[62] Figure 2(b)-(d) shows ARPES spectra of a 12 unit cell (approximately 12 nm) thick film of Na₃Bi thin film.[63] Figure 2(b) shows the observed band structure for the $\bar{\Gamma} - \bar{M}$ direction, while Figure 2(c) provides a representation of the 4D ARPES spectra in the $k_x - k_y - E_B$ 3D space around the centre of the BZ, where the intensity of the photoemission signal is rendered via the colour scale. Clear evidence of the formation of Dirac nodes can be seen in Fig. 2(c); visible about ±0.09Å from the Γ point along the $k_x$ direction are the two Dirac points [labelled D in Figure 2(c)]. The massless Dirac Fermion-like dispersion was further demonstrated by cuts taken along the $k_x$ directions at $k_y = 0$ for different values of $k_z$ shown in Figure 2(d): away from D, the bands appeared to show hyperbolic rather than parabolic dispersion, as expected for off-diameter vertical cuts of a conical surface. Linear and hyperbolic fitting of the ARPES data were used to extract the Fermi velocities, yielding $v_x = 4.2 \cdot 10^5\ ms^{-1}$ and $v_z = 1.3 \cdot 10^5\ ms^{-1}$. These values for the Fermi velocity, confirming the anisotropy of the dispersion, are in good agreement with several other reports on Na₃Bi bulk and thin films obtained by a variety of experimental techniques and substrates.[42,49,65,66,70,93]

The observation of a consistent valence band dispersion, independent of substrate, supports the idea of weak interaction between Na₃Bi and the substrates. Collins et al.[70] provided independent experimental evidence of this in the case of Na₃Bi on Si(111). Figure 2(e) shows the substrate Si*2p* core level before (black line) and after (red line) the growth of a few-layer Na₃Bi film. The spectral weight distribution of Si was found to be almost unaltered, with no new peaks arising and no evident broadening of the line-shape, indicating that no new chemical bonds were formed at the Si/Na₃Bi interface, meaning that Na₃Bi can be considered as freestanding on Si(111). The ARPES results in Figs. 2(b)-(d) indicate that Na₃Bi films are nearly charge neutral, with the Dirac point energy $E_D$ found to be 25 $meV$ above the Fermi energy, which we take to be the zero of energy, $E_F \equiv 0$, indicating p-type doping.

Na₃Bi thin films grown on sapphire were also found to possess minimal doping. Figure 2(f) shows scanning tunnelling spectroscopy (STS) of a 20 nm Na₃Bi film on Al₂O₃. The minimum in STS (black arrow) reflects the minimum density of states at the Dirac point energy, $E_D = -35\ meV$. Other experiments using ARPES or STS have found $E_D$ to range between $-50\ meV$ and $+25\ meV$,[94] depending on the substrate and on the presence of contaminants/vacancies on the surface created during growth. Similar growth-related self-doping effects altering the position of $E_D$ have been reported for tetradymite topological insulator materials: during the formation of Bi$_2$Se$_3$, selenium elemental vacancies contributing electron doping (n-type) are favoured;[95] in Sb$_2$Te$_3$ the Sb–Te defect usually leads to hole doping (p-type).[96] Multiple studies of Na$_3$Bi on Si(111)[60,61,63] or Al$_2$O$_3$[0001][59,60,66–69] have found p-type or n-type doping, respectively, and that exposure to Na overflux during a mild thermal annealing (200 ºC) post-growth can contribute to lowering the magnitude of $E_D$.

While there is an abundance of reports on the valence band structure of Na$_3$Bi thin films via ARPES,[42,63,70] studying the conduction band is more difficult, as photoemission cannot access the unoccupied bands above the Fermi level. While doping with electron donors (like alkali metals) can virtually move the band structure towards higher binding energy as a result of the extra charge, it can also result in an alteration of the electronic band structure due to the electric field from the residual positive ions on the surface,[97] preventing investigation of the band dispersion of the "pristine" crystal. Instead, a recent work by Di Bernardo et al.[61] shed light on the band structure well above the Fermi level by analysing quasi-particle interference (QPI) patterns, a method previously demonstrated to be effective for graphene,[98] superconductors[99] and other Dirac and Weyl semimetals, to probe the dispersion well above the Fermi level.[100]

Figure 2(g) shows a typical QPI pattern for a Na$_3$Bi film, produced by taking the Fourier transform of a large-area map of the differential conductance (d$I$/d$V$) at a constant tip bias voltage $V_b$. The ring-like structure in the QPI pattern is interpreted as due to the maximal scattering wavevector $q = 2k$, where $k$ is the in-plane electronic wavevector, at an energy $E = eV_b$ where $e$ is the elementary charge. Acquiring many maps at different energies thus allows mapping of the dispersions $E(k)$ of the Na$_3$Bi bands.

Figure 2(h) shows $E(k)$ for the 20 nm Na$_3$Bi film on Si(111) extracted from QPI (markers) vs. the predicted band structure calculated via HSE06 and GW methods. The experimental Fermi velocity extracted for electrons in the conduction band was $1.6x10^6\ ms^{-1}$, about 7 times higher than the velocity of holes. This is the same order of magnitude of graphene,[96] in good agreement with GW-based calculations [see Section 2 and Figure 2(h)] and previous reports on Na$_3$Bi ultrathin films,[70] while DFT predicts the velocity to be less than half this value. However, DFT methods yield better agreement with the valence band both in terms of the number of bands expected and their Fermi velocity. The origin of this discrepancy between GW and experiments is still unclear. One possibility is that the extra valence band observed experimentally originates from Na vacancies, creating an electric field on the surface of the sample which pushes the surface state up and away from the bulk states; however, this has been previously reported to result in a band with a much lower velocity. [49]

## 3.3. Semiclassical transport properties

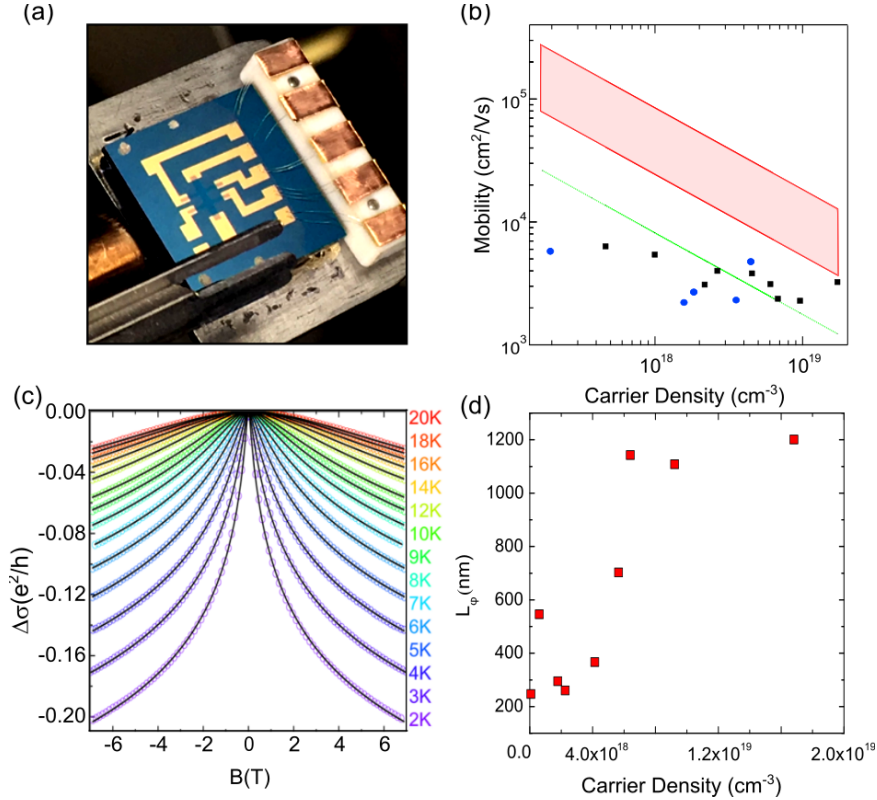

*Figure 3*: Semiclassical transport properties of Na$_3$Bi. (a) Picture of a Na$_3$Bi based Hall bar device on SiO$_2$. (b) Log plot of the experimental mobility vs. carrier density for 20nm Na$_3$Bi films grown Al$_2$O$_3$[0001] (data in black squares are extracted from Ref. [66]). The red shaded region is the maximum expected mobility assuming the only scattering is due to unipolar charged impurities, calculated using Eqn. 3 (Eqn. 18 in reference [101]). The width of the shaded region reflects uncertainty in the effective fine structure constant for Na$_3$Bi α = 0.06 – 0.17 (taken from Ref. [102]). (c) Change in longitudinal magneto-conductivity, $\Delta\sigma_{xx}$, as a function of a perpendicular magnetic field for a 20 nm-thick Na$_3$Bi layer capped with MgF$_2$, measured between 2 and 20K. Fitting with the HLN formula (solid line) is superimposed to the experimental data (open circles). (d) Coherence length against charge carrier density for 20 nm-thick Na$_3$Bi samples grown under different temperature conditions on Al$_2$O$_3$[0001]. Panels (b) and (d) adapted with permission from Hellerstedt, J. et al., Nano Lett. 16(5), 3210–3214 (2016). Copyright (2016) American Chemical Society. Panel (c) adapted with permission from Liu, C. et al., ACS Appl. Mater. Interfaces, 12 (31), 35542–35546 (2020). Copyright (2020) American Chemical Society.

The growth of Na$_3$Bi on insulating substrates such as sapphire and SiO$_2$ has enabled electronic transport measurements to be performed *in situ*, and the electronic properties to be probed as a function of growth conditions. Fuhrer and co-workers extensively investigated the electric properties of 20 nm thick Na$_3$Bi films.[58,59,66,68,69,103] For their transport measurements they grew the films on insulating, atomically flat Al$_2$O$_3$[0001] substrates in UHV via MBE which had been pre-patterned *ex situ* with Ti/Au electrodes evaporated through stencil masks to ensure a clean substrate surface. They were able to realize devices either in van der Pauw[66] or, by use of an insulating alumina stencil mask in contact with the substrate during Na$_3$Bi deposition, Hall bar geometries, like the one shown in Figure 3(a). These configurations allowed for the measurement of both the longitudinal and transverse magneto-resistance and magneto-conductance, $\rho_{xx}$, $\sigma_{xx}$ and $\rho_{xy}$, $\sigma_{xy}$ respectively, at a base temperature of 5 K and perpendicular magnetic field of up to 1 T. From these measured quantities, the Hall carrier density and mobility could be extracted: $n = \left(ed\frac{d\rho_{xy}}{dB}\right)$ and $\mu = (ne\rho_{xx})^{-1}$. Here, *d* is the film thickness.

Electron transport in a disordered Dirac semimetal has been modelled using the Random Phase Approximation (RPA).[101] In the homogeneous regime (carrier density $n$ greater than characteristic puddle carrier density $n^*$) the charge carrier mobility $\mu$ is given by

$$\mu = A\frac{e}{h}\frac{n^{1/3}}{n_i} \qquad (2)$$

where $e$ is the elemental charge, $h$ is Planck's constant, $n_i$ is the impurity density, and $A$ is a dimensionless constant which depends on the dimensionless interaction strength or fine structure constant $\alpha = e^2/\varepsilon\hbar v_F$, where ε is the relative dielectric constant, $\hbar$ the reduced Planck's constant, and $v_F$ the geometric mean of the Fermi velocities in the $k_x$, $k_y$, and $k_z$ directions. If the only disorder is due to dopant atoms, and all dopants are a single type, then the carrier density is equal to the dopant (impurity) density, $n = n_i$ and

$$\mu = A\frac{e}{h}n^{-2/3} \qquad (3)$$

As other types of disorder such as grain boundaries or (uncharged) point defects may contribute to scattering, and compensation of n-type and p-type dopants will lead to a number of impurities larger than the total carrier density ($n < n_{imp}$), this is an upper bound and we expect $\mu \leq A(e/h)n^{-2/3}$.

Figure 3(b) shows a log-log plot of the mobility against carrier density for 20 nm thin films of Na$_3$Bi on Al$_2$O$_3$[0001]: decreasing mobility with increasing carrier density was observed on both substrates, suggesting that dopants originating from impurities also acted as mobility-limiting scattering centres. The theoretical prediction in the red shaded region in Figure 3(b) was obtained using Eqn. 2 assuming a Fermi velocity for holes between $v_F = 1.4x10^5 ms^{-1}$ [66] and $v_x = 2.43x10^5 ms^{-1}$ [49] and a fine structure constant $\alpha = 0.06 - 0.17$. [66,102] The experimental data are seen to fall below the expected theoretical range, but approach it at the highest carrier densities. This likely indicates that other disorder, such as structural disorder (grain boundaries, point defects) is more important in limiting the mobility at low doping. These issues could be improved by further fine-tuning of the growth parameters, and by growing on substrates with better matched lattice constants. This would allow for higher quality films with less grain boundaries, larger islands and overall less disorder. If doping levels comparable to the lowest-doped films in Fig. 3(b) ($\approx 2x10^{17} cm^{-3}$) could be achieved with significantly lower structural disorder, then mobility in the range of $10^5 \ cm^2/Vs$ is conceivable.

## 3.4. Quantum transport properties

*3.4.1 Weak Anti-localization* At temperatures low enough for quantum coherence of electrons to be preserved over many scattering events, disordered 2D conductors typically show weak localization (WL), a quantum correction to the conductivity which manifests as a suppression of the conductivity at zero magnetic field due to constructive interference of backscattered electrons. A finite magnetic field destroys the constructive interference, and therefore WL results in a positive weak-field magnetoconductivity at low temperatures.

A striking aspect of Dirac semimetals is a π Berry phase for electron orbits which enclose the Dirac point. This Berry phase results instead in *destructive* interference of backscattered electrons and *enhanced* conductivity at zero magnetic field known as weak anti-localization (WAL), which leads to a *negative* weak-field magnetoconductivity. WAL therefore constitutes one of the hallmark transport signatures of topological Dirac semimetals.[104] It is fundamentally tied to the absence of backscattering of Dirac particles, and the impossibility of localizing a massless particle due to Klein tunnelling.[105] WL and WAL manifest themselves most strikingly in 2D through a logarithmic temperature dependence, hence thin films are best suited to their observation (film thickness *t* less than the phase coherence length $l_\phi$). Thus, thin epitaxial films of Na$_3$Bi open the possibility to study this quantum-coherent transport phenomenon, probing the Berry phase and the phase coherence length.

The magnetoconductivity due to WAL can be described by the Hikami-Larkin-Nagaoka (HLN) formula in the limit of perfect spin-orbit coupling[106]:

$$\sigma(B) - \sigma(0) = -\frac{\alpha e^2}{2\pi h}\left[ln\left(\frac{B_\phi}{B}\right) - \Psi\left(\frac{1}{2} + \frac{B_\phi}{B}\right)\right] \quad (4)$$

Here, the coherence field is $B_\phi = h/8\pi e l_\phi^2$, $\alpha$ is a parameter indicating the type of localization ($\alpha = \frac{1}{2}$ for WAL), and Ψ is the digamma function. An even more detailed model to describe the quantum correction to the magnetoconductivity of topological insulator surface states and Weyl semimetal films, taking into account extrinsic spin-orbit scattering, was developed by Adroguer et al.[107] The correction to the magnetoconductivity was found to be always positive, predicting weak antilocalization in either the presence or absence of spin-orbit impurity scattering.

Figure 3(c) plots the typical magnetoresistance for Na$_3$Bi thin films grown on sapphire and capped against air exposure (see Section 5.2.1) demonstrating the negative magnetoconductance expected for WAL. A fit to Eqn. 3 is shown as solid black lines, yielding $\alpha \in [0.3; 0.5]$ while $l_\phi$ exhibits a power law dependence on temperature. The excellent agreement with Eqn. 3 indicates that the description of quasiparticles in Na$_3$Bi as Dirac electrons is excellent; in particular inter-valley scattering (scattering between Dirac nodes) which would destroy WAL, is weak. From similar magnetoresistance measurements, Hellerstedt et al.[66] were able to extract the coherence length as a function of the carrier density at a temperature *T* = 5 K. The results in Figure 3(d) are for the same samples denoted by black dots in panel (b). The measured phase coherence length is always larger than the nominal thickness of 20 nm, confirming that the devices are 2D in terms of quantum coherence, and can be as large as 1 μm. Even larger phase coherence lengths are anticipated at lower temperatures. The long phase coherence length implies that mesoscopic devices (in which device dimensions are comparable to the phase coherence length) could be fabricated with conventional micro- or nano-fabrication techniques. This paves the way for further exploration of quantum coherent phenomena in Na$_3$Bi including Fabry-Perot interferometry,[102] Klein tunnelling, the Aharonov-Bohm effect, and others, as observed for $Cd_3As_2$,[108–111] involving both bulk and surface Fermi arc states.

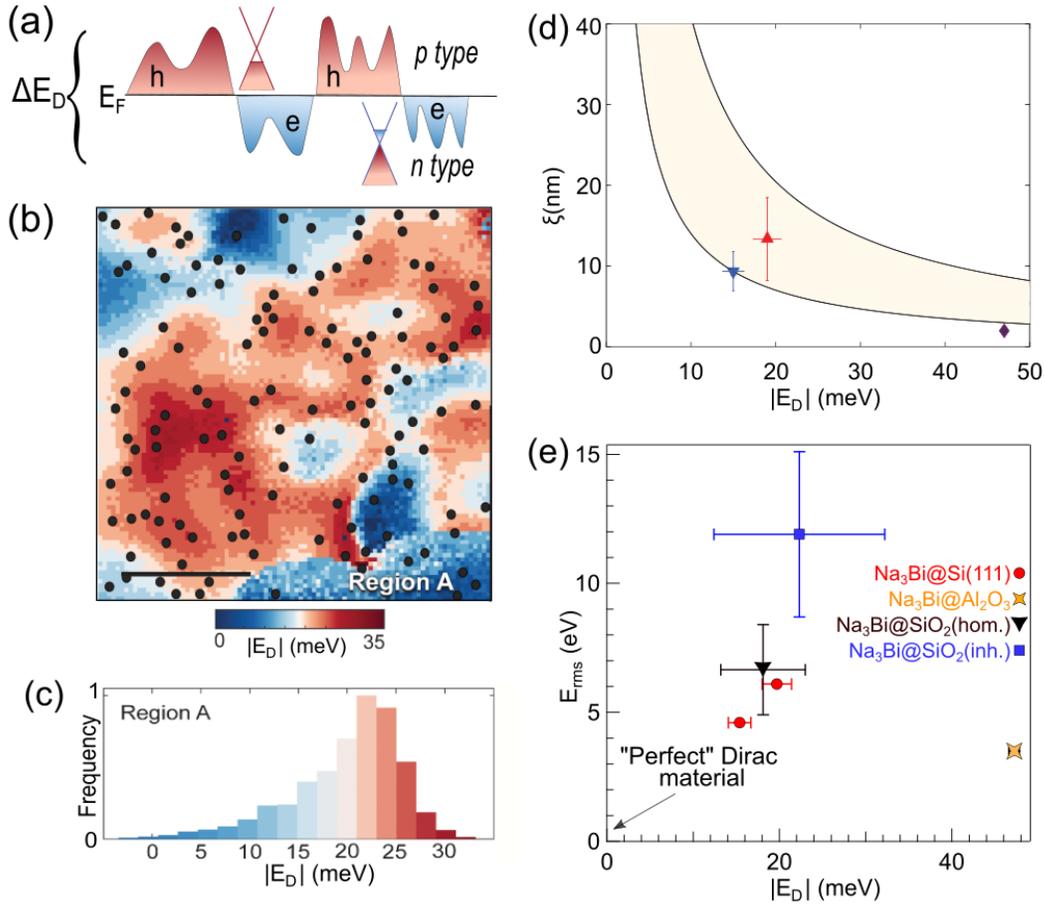

*Figure 4:* Charge puddling in $Na_3Bi$ at low carrier density. (a) Schematic of charge puddles. (b)Dirac point energy map of a (45 nm x 45 nm) region of p-type $Na_3Bi(001)/SiO_2$; scalebar, 15nm. (c) Histogram of the Dirac point energy from (b), color-coded to reflect the intensity in map (a). Black points in (b) show the location of Na vacancies.[60] (d) Spatial coherence length ξ for charge puddling as a function of Dirac point energy for three different samples. Shaded area indicates the expected Dirac behavior $\xi \sim E_D^{-1}$ with the width of the shaded area reflecting uncertainty in the materials parameters for $Na_3Bi$. (e) Plot of the potential fluctuations in $Na_3B$ films due to charge puddling as a function of the Dirac point energy, for films on various substrates as indicated in the legend.[60,103] Panels (b)-(d) adapted from Edmonds, M.T, et al., Sci. Adv. 2017, 3, eaao6661.

*3.4.2 Accessing Dirac-point physics, and electron-hole "puddling"*

When Dirac materials are doped far from charge neutrality, they have a large Fermi surface whose properties are in many ways similar to that of non-topological materials. The differences become more pronounced close to the Dirac point where the Fermi surface and metallic carrier density shrinks to zero. When topological materials are doped close to the Dirac point, new physics starts to play an increasingly important role including: (i) Thermally excited electron-hole pairs dominate over the intrinsic carriers, creating a unique "Dirac plasma" state of matter at finite temperatures. Provided the Dirac plasma is sufficiently clean, it is believed to be a hydrodynamic fluid whose properties are governed by a non-Fermi liquid quantum-critical conformal field theory.[112,113] (ii) Even at low temperature, Coulomb interaction effects are still expected to play a critical role, for example in monolayer graphene, the long-range Coulomb interaction strongly renormalizes the electron velocity while a short-range Hubbard U can break time-reversal symmetry destroying the topologically protected Dirac points;[114] (iii) A magnetic field creates unique chiral modes which disperse only along the field direction. Observation of these phenomena requires close proximity to the Dirac point: $|E_D| < k_B T$ (Boltzmann constant) and $|E_D| < \hbar\omega_c$ (cyclotron frequency) are the characteristic scales defining the Dirac plasma regime and the ultra-quantum limit in which the only populated states are chiral Landau levels.[113]

Close approach to the Dirac point is governed by the unique interplay of disorder and screening in a Dirac semimetal. Far from the charge neutrality point, the carrier density is homogeneous i.e. spatial fluctuations in the carrier density are smaller than the average carrier density and disorder can be incorporated into a translationally invariant theory as a disorder-broadening of the quasiparticle Green's functions. However, as the Fermi energy approaches the Dirac point, the ability of Dirac electrons to screen the disorder potential vanishes, and the screening length, proportional to the Fermi wavelength (the only length scale in a Dirac system) diverges. As a result, it becomes more energetically favourable for the system to break up into electron-like and hole-like spatial regions. Figure 4(a) illustrates this schematically; the system breaks translation invariance, and a semi-classical approach is required where the dispersion remains unchanged by disorder, but shifts up or down in energy following the impurity potential. The details of puddle formation depend sensitively on the screening properties: the unscreened impurity potential itself induces carriers (either electrons or holes) which then screen the potential. Thus, the screened potential, and the local carrier density, of a disordered Dirac electronic system must be solved self-consistently. This approach was first developed for graphene,[115] and since modified to encompass 3D TDS.[116]

The spatial fluctuations in Figure 4(a) can be characterized by an $rms$ carrier density $n_{rms}$. This residual carrier density leads to a finite conductivity even when the average chemical potential is at $E_D$. This residual minimum conductivity is $\sigma_{min} \approx n_{rms} \mu \approx \frac{n_{rms}}{n_i}$, as by Matthiessen's rule, $\mu \approx 1/n_i$. For monolayer graphene, it was found [Ref. [109]] that the ratio $\frac{n_{rms}}{n_i}$ was mostly constant (it had a weak dependence on $\log(\sqrt{n_i}\, a)$, where $a$ is the distance between the impurities and the 2D graphene plane). Since the conductivity also scales as the ratio $\frac{n_{rms}}{n_i}$, these electron-hole puddles explained one of the most striking early observations about graphene: why different samples all showed a "quasi-universal" conductivity at charge neutrality. However, the conductivity was not universal: as predicted, as improvements in sample quality gave orders-of-magnitude reduction in $n_i$, there was a corresponding linear increase in the minimum conductivity as $n_i$ was reduced exponentially.[117]

The spatial variation of the charge density, reflecting the charge screening length, also shows unique behaviour in a Dirac system. The only characteristic length scale for a Dirac electronic system is the Fermi wavelength $\lambda_F \approx k_F^{-1} \approx |E_F - E_D|^{-1}$, where $k_F$ is the Fermi wavevector. This is in sharp contrast to a massive system: for example, in a 2D massive system $k_F^{-1} \approx a_B$, the Bohr radius. As $|E_F - E_D| \to 0$ we expect the screening length of a Dirac system to diverge. This is cut off by the same puddling behaviour described above; the only remaining length scale is $n_i^{-1/2}$. Indeed, the correlation length $\xi$ for charge fluctuations in graphene at $E_F = E_D$ behaves as $\xi \approx n_i^{-1/2}$ ($\xi$ is defined as the FWHM of the autocorrelation function $\langle v_{sc}(r)v_{sc}(0)\rangle$ for the screened Coulomb potential). Far from $E_F = E_D$ the self-consistent fluctuations are less important, and $\xi \approx k_F^{-1} \approx |E_D|^{-1}$.

The self-consistent theory for Na$_3$Bi was worked out in Ref [60]. Here it was assumed that the impurity potential comes from a random three-dimensional distribution of charged impurities with density $n_i$. Within the Thomas-Fermi approximation, it can be shown that at charge neutrality, just like monolayer graphene, $\frac{n_{rms}}{n_i}$ is constant. Similarly, the correlation length shows Dirac behaviour, with $\xi \approx k_F^{-1} \approx |E_D|^{-1}$ for large doping $n \gg n_{rms}$ and $\xi \approx n_i^{-1/3}$ as $|E_D| \to 0$ reflecting that there is no other length scale present.

The spatial variation of the local chemical potential in Na$_3$Bi due to disorder near the Dirac point has been probed directly in experiments. Figure 4(b) shows the spatial variation of the Dirac point energy $E_D$ measured via scanning tunnelling spectroscopy on a 20 nm thick epitaxial film of Na$_3$Bi on Si(111). The local $E_D$ is indicated using a colour scale, and a histogram of $E_D$ is shown in Figure 4(c). The $rms$ width of the $E_D$ distribution is $E_{rms} = 5.6\ meV$. This is comparable to the spatial variation of Dirac point energies (5.4 $meV$ in Ref. [118]) observed for graphene on hexagonal boron nitride (h-BN). Graphene on h-BN is known to have exceptionally low variation in the Dirac point energy,[118] which has enabled numerous experimental probes of Dirac point physics.

From the measurement in Figures 4(b) and 4(c), an impurity density of $n_i = 3 - 5 x10^{18} cm^{-3}$ can be inferred using the self-consistent theory, which would in turn imply a Coulomb impurity-limited charge carrier mobility of $6,000 - 17,000\ cm^2/Vs$. While transport measurements could not be performed on this film grown on conducting Si(111), similar measurements were used to infer $n_i = 2 - 4\ x10^{18} cm^{-3}$ and $\mu = 19,000 - 52,000\ cm^2/Vs$ for a film on insulating Al$_2$O$_3$ for which transport measurements could also be performed. Indeed, Hall effect measurements confirmed a carrier density of $4.35\ x10^{18} cm^{-3}$, indicating that the dopants are charged impurities. However, the Hall mobility was only $3,450\ cm^2/Vs$, indicating that other disorder plays a role in limiting the mobility, consistent with the discussion in Section 3.3.

Figure 4(d) shows the correlation length $\xi$ determined from the autocorrelation of measurements such as Fig. 4(a), for three different samples. The shaded region represents the expected Dirac behavior at high doping $\xi \approx k_F^{-1} \approx |E_D|^{-1}$, where the constant of proportionality is determined only by Na$_3$Bi materials parameters fine structure constant $\alpha$ and the average velocity $v_F$ near the Dirac point. Since $\alpha$ and $v_F$ are not known precisely, a range is used to represent the best experimentally obtained values, with the upper bound using $v_F = 2.4\ x10^5 ms^{-1}$ and $\alpha = 0.069$ and the lower bound using $v_F = 1.4\ x10^5 ms^{-1}$ and $\alpha = 0.174$. The experimentally measured coherence length agrees well in magnitude and in Fermi energy dependence with the self-consistent theory for a 3D Dirac semimetal, and the better match with the lower bound is more consistent with stronger interactions (larger $\alpha$).

Growth of Na$_3$Bi on SiO$_2$/Si substrates has allowed electronic transport measurements incorporating a back-gate to tune the carrier density. Damage to the SiO$_2$ due to exposure to Na during growth limits the effectiveness of SiO$_2$ as a gate dielectric, however a change of carrier density $\approx 10^{18} cm^{-3}$ could still be achieved. In combination with charge transfer doping from F4-TCNQ (2,3,5,6-Tetrafluoro-7,7,8,8-tetracyanoquinodimethane), this enabled tuning from the homogeneous electron-doped regime $n > n^*$ to the inhomogeneous regime $n < n^*$. Similar to gate-tuned graphene, the conductivity exhibited a broad minimum in the inhomogeneous regime. The variation in both the longitudinal and Hall conductivities across this transition could be modelled well using the self-consistent transport theory and effective medium theory, as employed successfully to understand transport in gated graphene. However, unlike graphene, accurate modelling of the behaviour of Na$_3$Bi requires accounting for different electron and hole mobilities, and in fact the hole mobility was found to be about ten times higher than the electron mobility. This result is somewhat unusual, as typically semiconductors have higher mobility for electrons than holes (due to smaller effective mass for electrons), however the higher hole mobility follows naturally from the inverted band structure of Na$_3$Bi which leads to a lower velocity for holes compared to electrons.

Figure 4(e) summarises $E_D$ and $E_{rms}$ achieved for Na$_3$Bi on various substrates. The small $E_D$ and $E_{rms}$ values for Na$_3$Bi suggest that thermal activation will be important even at moderate temperatures, as the characteristic values of $E_D$ and $E_{rms}$ in Figure 4(c) are lower than the thermal energy $E_D$ and $k_B T$ at room temperature. Indeed, temperature-dependent transport on Na$_3$Bi films on Al$_2$O$_3$ has shown strong variation in the longitudinal and Hall resistivities with temperature, including a sign change in the Hall effect.[68] These effects could be well modelled by considering the effect of thermal activation near the Dirac point. It was found that the hole mobility was about seven times larger than the electron mobility, independently confirming the similar result from gated transport measurements. Temperature-dependent transport also indicates a strong dependence of the mobility on temperature, suggesting that intrinsic electron-phonon scattering may limit the mobility to only $\approx 300\ cm^2/Vs$ at room temperature.

## 4. 2D confinement

Topological Dirac semimetals (TDS) combine aspects of topological and trivial insulators within their band structure. The Dirac points reflect a change in topology as a function of momentum parallel to the direction connecting them; the bands are inverted (topological) between the Dirac points, and non-inverted (trivial) outside the Dirac points. This transition from topological to trivial within momentum space is reflected in the Fermi-arc surface states, remnants of the topological surface states of a 3D topological insulator, terminating at momenta which reflect the topological-trivial transition in momentum space at the Dirac points.

The Dirac points of a TDS lie at finite momenta (wavevector), not necessarily commensurate with the lattice periodicity. This offers opportunities to reveal and control the hidden momentum-space topology by confinement, which selects certain momenta by quantizing the wavevector parallel to the confinement direction, due to the particle-in-a-box confinement effect. In TDS (and similarly in 3D topological insulators) reduction to thin and ultrathin films is expected to induce the formation of a bandgap, whose size and topological nature depends on the confinement direction and film thickness.[119–121]

In ultra-thin form, Na$_3$Bi has been predicted to be a quantum spin-Hall insulator, or a 2D topological insulator, which is a time-reversal invariant phase of matter that exhibits a 2D bulk bandgap while hosting metallic edges.[5,122] The edge states in a QSH insulator are helical, with spin-orbit coupling (SOC) forcing electrons of opposite spins to travel in opposite directions. Backscattering is prevented by time-reversal symmetry: the QSH edges are expected to be perfect conductors at zero temperature in the absence of magnetic disorder. These helical edge modes offer the possibility of ballistic spin-polarized edge currents, as demonstrated experimentally in HgTe quantum wells,[123] a feature that could be harnessed for the realization of dissipationless transport devices and low-power information processing.

In this section we give an overview of the effects of quantum confinement on Na$_3$Bi from theoretical and experimental perspectives, specifically considering the thickness dependence of the bandgap, the Stark-effect induced topological phase transition in the ultrathin films, and the emergence of giant negative magnetoresistance phenomena.

## 4.1. Thickness-dependence of bandgap

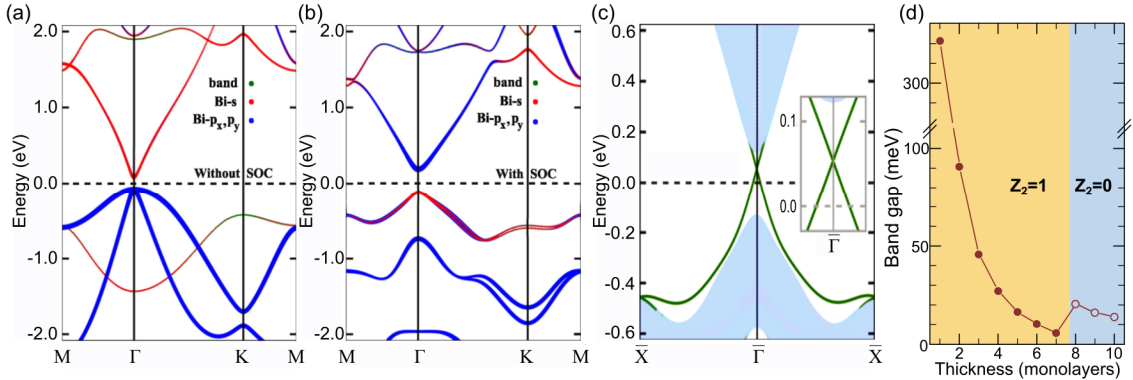

*Figure 5:* Theoretical description of the dependence of the bandgap of Na$_3$Bi on the thickness of the film. (a) and (b) Orbitally-resolved PBE calculations for the band structure of Na$_3$Bi ML without and with spin-orbit coupling, respectively. (c) Localization resolved edge states of ML Na$_3$Bi, the inset shows the zoom-in at the $\bar{\Gamma}$ point. The colorscale from light blue to green represents the weight of atoms located from the middle to one edge of the ribbon. (d) Calculated bandgap and $\mathbb{Z}_2$ values as a function of film thickness. Panels (a)-(c) adapted with permission from Niu, C. et al., Phys. Rev. B 95, 075404 (2017). Copyright (2017) by the American Physical Society. Panel (d) Adapted with permission from Xia H. et al., ACS Nano 13, 9647–9654 (2019). Copyright (2019) American Chemical Society.

Anisotropic confinement in Na$_3$Bi has been initially explored by Xiao et al.[71]: they found that confinement in the horizontal direction (i.e., along the $a − b$ plane of the crystal structure) splits the continuum energy spectrum of Figure 1(d) into subbands, and opens a finite bulk bandgap at the Dirac point of width inversely proportional to the ribbon size. Furthermore, they predicted that because of inversion symmetry, there exist gapless topological surface states, protected by TR from opening a gap.

Simple consideration of the bandstructure of Na$_3$Bi, which shows a trivial gap at large c-axis momenta, indicates that c-axis oriented films in the ultrathin limit should be conventional insulators. Indeed, it was predicted theoretically based on a simple tight-binding model of Na$_3$Bi that films thinner than $t = \pi/k_D$, where $k_D$ is the Dirac point momentum, should be trivial insulators.[71] However more recent first-principles calculations by Niu et al.,[124] later confirmed by Yang and coworkers,[70] predicted that monolayer (ML) Na$_3$Bi (as well as bilayer Na$_3$Bi[70]) is a topological insulator. Figure 5(a) and 5(b) report the orbital-resolved calculated band structure for a ML of Na$_3$Bi without and with the inclusion of spin-orbit coupling, respectively, calculated via the PBE method. Switching on SOC induces a *s-p* band inversion at the Γ point, with the system preserving its insulating nature but increasing the bandgap from 160 to 310 meV. The authors confirmed the topological nature of this electronic system [see Figure 5(c)] by investigating the edge states of 1D nanoribbons of the Na$_3$Bi ML: Kramers edge states (green) were clearly observed within the bulk bandgap, forming linearly-dispersing Dirac cones with the apex touching at the Γ point. In the work by Collins et al.[70] detailed calculations showing similar results for a Na$_3$Bi bilayer are also reported. The topological invariant of the edge states was further confirmed using the Wilson loop method,[125,126] by tracing the evolution of the Wannier function centres.

A topologically non-trivial scenario was subsequently predicted by Xia et al.[65] for films with thickness up to 7 ML; for thicker films interlayer coupling is expected to induce a further $s - p$ band inversion and a transition to a topologically trivial state. Figure 5(d) shows the size of the bulk bandgap reported by Xia et al. for Na$_3$Bi as a function of layer thickness, calculated with a PBE XC functional. The exact bandgaps for mono- and bilayer differ from the work of Yang's group,[70] probably due to a different approach to the XC potential used in the calculations, but the 7-layer periodicity in the topological nature of the electronic system presented in Ref. [71] was confirmed, as well as the presence of non-trivial topological states in the first 7 layers. The gap was calculated to be 46 (27) $meV$ for 3 (4) MLs; the experimental values were found to be slightly larger (see discussion below).

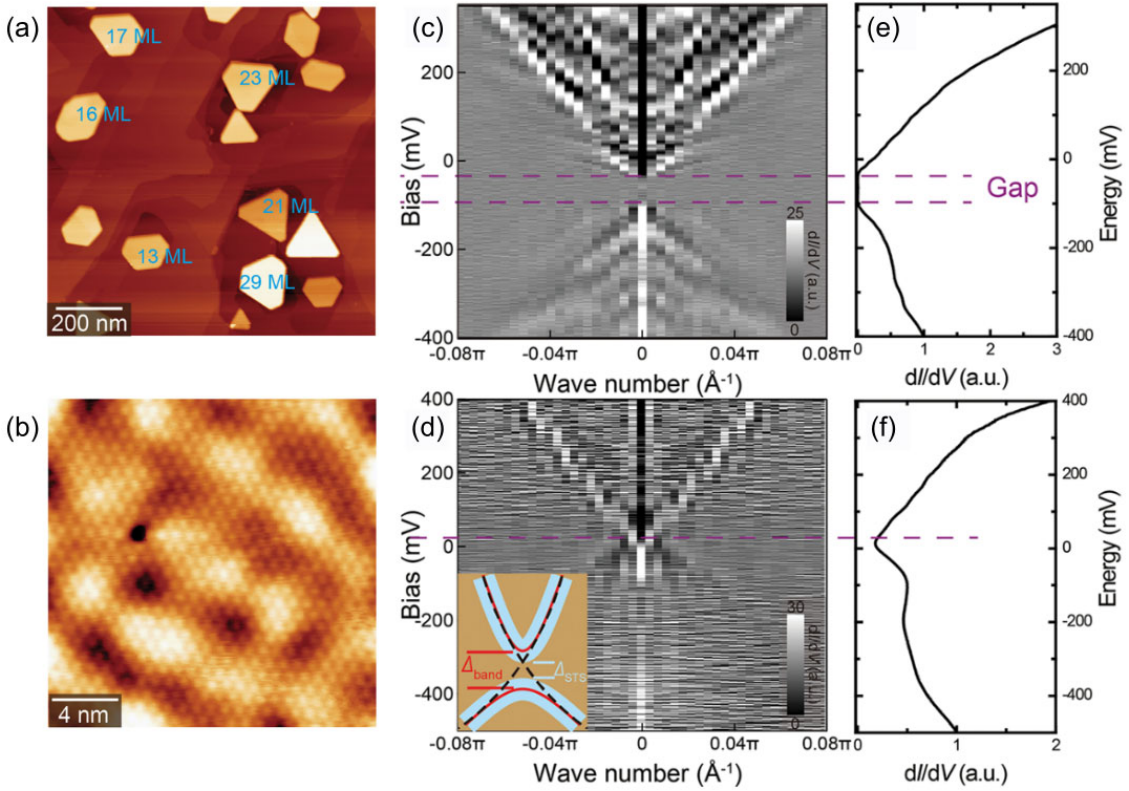

**Figure 6:** *Bandgap dependence of $Na_3Bi$ on film thickness. (a) STM image of $Na_3Bi$/graphene (bias voltage V = 3.0 V and tunnel current I = 10 pA), showing islands of different thicknesses. (b) Atomic resolution of a $Na_3Bi$ film grown by optimizing the Na flux (V =0.1 V and tunnel current I = 100 pA), (c) Second derivative of the 1D FFT of a standing wave pattern obtained by acquiring dI/dV spectra on a 3ML island of $Na_3Bi$, to show the energy-dispersion relation. (e) Average STS on a 3ML island. (d) and (f) are the same as (c) and (d) on a 5ML island. Purple horizontal dashed lines indicated gap edges. Inset of panel (d) is a sketch for bandgap size determination. Adapted with permission from Xia H. et al., ACS Nano 13, 9647–9654 (2019). Copyright (2019) American Chemical Society.*

Experimental evidence of the bandgap dependence on the thickness of $Na_3Bi$ was first provided by Collins et al.[70] and then further examined by Xia et al.[65] for ultra-thin $Na_3Bi$ grown on Si(111) and bilayer graphene (BLG), respectively. Figure 6(a) reports a large-scale topographic image of an as-grown $Na_3Bi$ film grown by Xia et al.[65] on graphene-covered SiC, showing hexagonal-shaped islands of different thicknesses (atomic resolution in Figure 6(b)). The authors examined the thickness dependence on the bandgap both by performing scanning tunnelling spectroscopy (see Figure 7(c) below) and by analysing quasiparticle interference patterns. By acquiring 2D conductance mappings near the step edges of 3ML and 5ML islands, they observe standing wave patterns generated by quasiparticles scattering against the step edge. 1D-FFT analysis of such patterns reveals linearly dispersing branches, corresponding to the scattering modes of Dirac bands in $Na_3Bi$: while the 3ML film shows a clear bandgap of ~70 meV, consistent with the average STS spectra shown in Figure 6(e), the 5ML film only exhibits a spectroscopic dip around -10 meV, and as shown in Figure 6(f) the observed conductance is always finite. The discrepancy between the experimental and the predicted bandgap was attributed to the approximations intrinsic to the calculations and to experimental limitations: the small bandgap predicted for thicker layers is challenging to confirm experimentally, given the finite resolution of the available spectroscopies and the inevitable disorder-induced spectral broadening in the system.

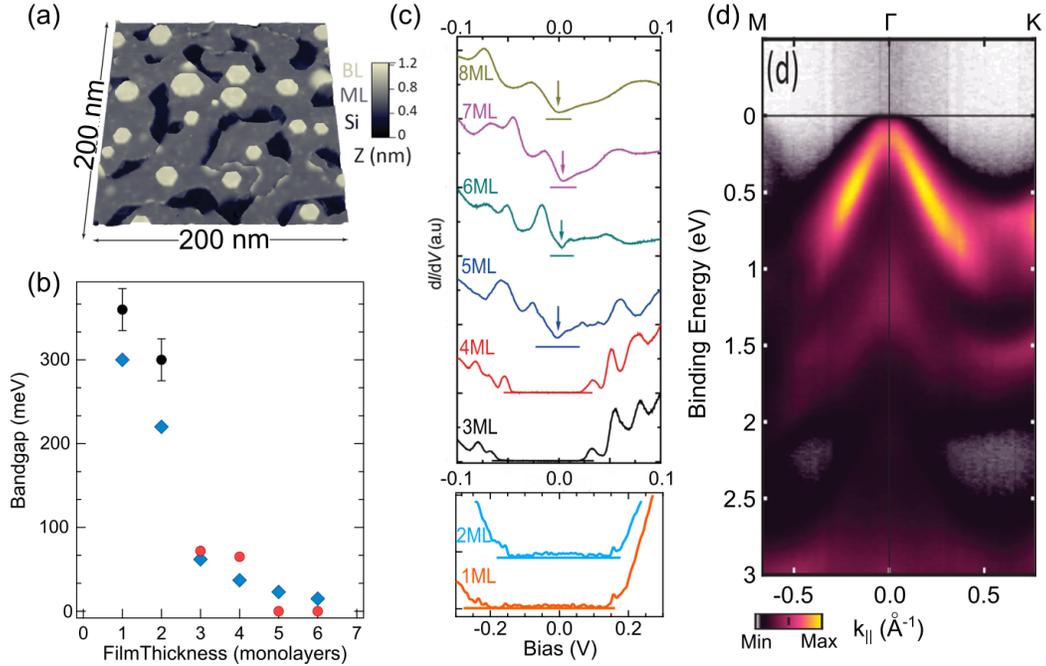

*Figure 7: Experimental evidence of bandgap dependence on film thickness. (a) (200 nm x200 nm) topographic STM image of ML/BL of $Na_3Bi/Si(111)$ (bias voltage V = 2.0 V and tunnel current I = 120 pA).[70] (b) Comparison between the predicted (blue diamonds) and measured bandgap of $Na_3Bi$ as a function of layer thickness. Black dots and red dots are extracted from Ref. [70] and Ref.[65], respectively. (c) dI/dV spectra acquired on regions of $Na_3Bi$ with different thicknesses; spectra offset vertically for clarity. Horizontal bars mark the zero differential conductance for each curve. Top and bottom panels adapted from Ref.[65] and Ref. [70], respectively. Note different horizontal scales in top and bottom panels. (d) Angle resolved photoemission spectra of the sample in (a) along the $\bar{M} - \bar{\Gamma} - \bar{K}$ direction; photon energy $h\nu = 48\ eV$. Data in (b) and (c) Adapted with permission from Xia H. et al., ACS Nano 13, 9647–9654 (2019). Copyright (2019) American Chemical Society. Panels (b) and (d) adapted from Collins, J., et al., Nature 2018, 564, 390.*

Figure 7 summarizes the results of Collins et al.[70] and Xia et al.[65]. A large ($200\ nm\ x\ 200\ nm$) topographic image of MBE-grown ML/BL $Na_3Bi/Si(111)$ is presented in panel (a), where regions are colour coded according to the number of layers: white for bilayer (BL), dark grey for ML, and dark blue for underlying Si. Monolayer regions were identified by an additional 2.2 Å distance from the substrate due to structural relaxation.[70] In both works, atomically flat islands up to 40 nm in lateral extent and were observed, allowing for spectroscopy on areas of different thickness. The extracted bandgap for ML and BL is 0.36 and 0.30 eV respectively, larger than thermal excitations at room temperature. The bandgap size for up to 6ML is reported in Figure 7(b), along with the theoretical prediction according to PBE calculations (blue diamonds), exhibiting the predicted trend, decaying roughly exponentially with thickness. The STS spectra from Collins et al.[70] and Xia et al.[65], which reflect the energy dependent LDOS for different layer thicknesses, are presented in Figure 7(c); here the spectra are offset vertically for clarity and the color-coded horizontal bar represents the zero of the differential conductance for each spectrum. A gap of progressively smaller size (note the different horizontal scales in the top and bottom panels) is observed up to 4ML, while finite conductance was measured by Xia et al. for thicker samples, with only a dip reported in the DOS and an eventual evolution towards the bulk limit after 8ML.

The existence of a large bandgap can also be seen directly in ARPES. Figure 7(d) shows the valence band structure of 1-2 ML Na$_3$Bi acquired along the $\bar{M} - \bar{\Gamma} - \bar{K}$ direction by Collins et al.[70] A valence band maximum is observed at $\bar{\Gamma}$, with maximum 0.12 eV below the Fermi energy which lies in the bandgap. Three subbands peaked at $\bar{\Gamma}$ are visible in Fig. 7(d), consistent with expectations for 1 ML and 2 ML Na$_3$Bi. The broadness of the spectroscopic features was attributed to the coexistence of the two thicknesses, which have slightly different band structures, and to the relatively small crystal size of each domain.

## 4.2. Topological phase transition with electric field

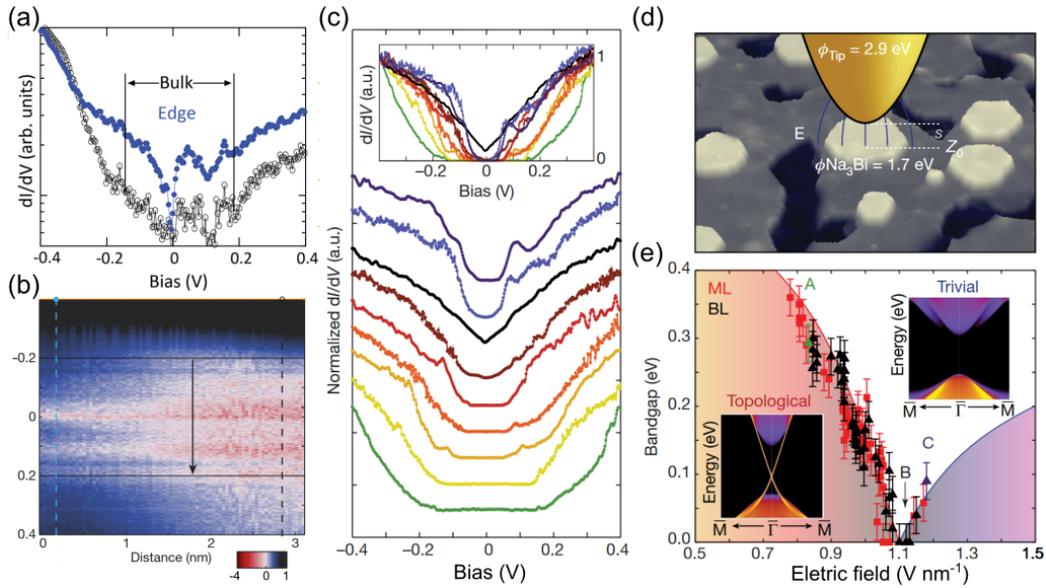

*Figure 8:* Topological to trivial transition of edge states. (a) Log-scale plot of differential conductance spectra acquired on a step edge of BL Na$_3$Bi (blue) and on the bulk of the BL (black). (b) dI/dV colour map taken at the edge (blue dashed line on the left) of Na$_3$Bi and moving away from it. Colour scale shown at the bottom right. (c) dI/dV spectra acquired on BL Na$_3$Bi for different tip-sample separations (different electric fields); spectra are offset for clarity and overlayed in the inset. (d) Schematic of a typical STM/STS measurement at tip-distance separation S, where the difference in work function between tip and sample generates an electric field. (e) Bandgap extracted from STS for ML (red triangles) and BL (black squares) as a function of the applied electric field. Adapted from Collins, J., et al., Nature 2018, 564, 390.

Edmonds and coworkers[70] also directly observed the topological edge state within the bulk gap of BL Na$_3$Bi via STS. Fig 7(a) reports the comparison between a differential conductance spectrum acquired on the edge of a BL Na$_3$Bi island on Si(111) (blue curve) and one acquired 3 nm inwards away from the edge. In the 2D bulk (black curve) a clear bandgap can be seen, whereas on the edge finite conductance is always observed with a characteristic intensity dip at $V_b = 0$ (the Fermi level). The authors demonstrated the extended nature of this edge feature by measuring STS spectra as a function of the distance from the edge, as in Figure 7(b). Here, the intensity is represented via the colour scale, where an increasingly white/red colour corresponds to a decrease in the signal. In their work, by integrating the signal along the columns (along the direction of the black arrow) in the area enclosed by horizontal black lines, they were able to observe the exponential decay in intensity (not shown) expected for a 1D topologically non-trivial state.[127]

As early as 2015 Yang and coworkers[128] proposed the possibility to induce a topological transition in ultra-thin TDS $A_3Bi$ (A= Na, K, Rb) and Cd$_2$As$_3$ by means of applying an electric field, developing on the simple model discussed in Section 2. By assuming a relatively small potential $V = eEz$ along the *c* axis, and treating it as a perturbation, it was demonstrated that a Stark-effect driven band inversion at the Γ point would occur in a film thin enough to meet the quantum confinement criteria.

This prediction was verified experimentally by Collins et al.[70] on ML and BL Na$_3$Bi by means of STS measurements. When in tunnelling range, the difference in work function between the metallic tip and sample $\phi_{Tip} - \phi_{Na_3Bi}$ generates a localized electric field *E*. By tuning the tip-sample separation, it is therefore possible to tune the electric field applied to the sample. Spectra collected at distances ranging from $1.02\ nm$ (purple spectrum, top) to $1.45\ nm$ (green spectrum, bottom), corresponding to an applied *E* of $1.18\ nm^{-1}$ to $0.83\ V\ nm^{-1}$, respectively, are shown in Figure 7(c). A model of the physical setup is portrayed in Figure 7(d). As the electric field strength is increased (bottom to top), the bandgap was observed to go from a width of approximately $300\ meV$, to be fully closed at $1.1\ V\ nm^{-1}$ (black spectrum), and then reopen with a further increase of electric field, reaching 90 meV for an electric field of $1.2\ V\ nm^{-1}$. In the inset of Figure 7(c), where the spectra are overlayed without vertical stacking, the authors point out that the black spectrum exhibits the typical V-shape of a TDS, where the conductance has a dip corresponding to the Dirac point but it is always finite. The value of the bandgap measured for ML and BL as a function of the applied electric field is presented in Figure 7(e): in both cases a critical field was observed, where the bandgap is closed and then reopened in the conventional regime. Projected edge state band structures below and above the critical field (as calculated by DFT for a Na$_3$Bi ML along the $\bar{M} - \bar{\Gamma} - \bar{M}$ direction) predicting the topological to trivial transition are reported as insets in Figure 7(d). In these simulations, the critical field for which the bandgap is observed to close and then reopens is about $1.85\ V\ Å^{-1}$; the discrepancy with the experimental value has been attributed to the presence of Na vacancies in the simulated supercell, which can screen external electric fields and cause an overestimation of the critical field.

## 4.3. Edge-state transport, giant negative magnetoresistance

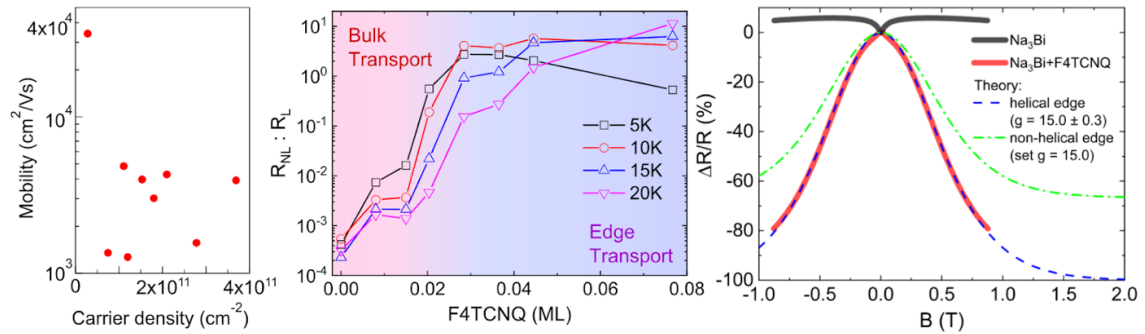

*Figure 9: Edge state transport in Na$_3$Bi ultrathin films. (a) Mobility against carrier density at 5.3 K for 2-nm Na$_3$Bi grown on Al$_2$O$_3$[0001]. (b) Ratio of the non-local to local resistance as a function of deposited F4-TCNQ molecules (so as a function of*

*increasing p doping). (c) Resistance as a function of magnetic field at T = 5.3 K for bulk-conducting film (as grown, black line) and edge-conducting film (0.015 ML F4-TCNQ coverage, red line). Fits to theory for exchange-mediated scattering for helical and non-helical edges are shown as blue dashed and green dot-dashed lines, respectively. Adapted from Ref.[59].*

Spin-polarized edge transport is one of the defining properties of 2D topological insulators, and perfect ballistic conduction is expected in the QSH state at 0 K. Experimental evidence of 1D edge transport has been reported for very few systems. Quantised ballistic transport has been observed over few-micron distances in HgTe quantum wells[129,130] and ~100 nm in WTe$_2$.[131] Non-local resistance measurements have also indicated edge transport over larger length scales in HgTe,[132] WTe$_2$,[133] and the GaSb/InAs interface.[134] However edge transport alone does not guarantee a topological insulator, and indeed edge transport has been observed to persist in some cases in regimes where it is not expected by topological considerations.[135] Directly demonstrating the helical nature of topological edge transport has been challenging. Spin-polarized transport has been observed for the edges of HgTe quantum wells, however the experiments were semi-quantitative, and could not demonstrate perfect spin polarization.[123]

Ultrathin films of Na$_3$Bi provide a new system to study topological edge transport. These films are particularly attractive due to the very large bandgap ($> 300\ meV$,[70] much greater than $k_BT$ at room temperature) and the persistence of the topological insulator state with a large bandgap up to a thickness of several layers, which indicates that Na$_3$Bi is relatively robust to layer-number fluctuations that might be caused by imperfect growth. Recently ultrathin films of Na$_3$Bi have shown clear evidence of edge transport over millimetre distances through non-local measurements.[59] The simplicity of Na$_3$Bi's structure, with a well-defined spin quantization axis out of the plane of the film, allows a unique electrical transport signature of helical edge transport in the form of a giant negative magnetoresistance, where the magnitude of the magnetoresistance can be used as an unambiguous probe of helicity.

Liu et al.[58,59] were able to prepare ultrathin Na$_3$Bi of 1-3 ML thickness by MBE on insulating, flat Al$_2$O$_3$[0001] prepatterned with electrodes, enabling in situ electronic transport measurements in UHV. Figure 9(a) reports the observed mobility versus carrier density for several as-grown 1-3 ML thick films: all films exhibited majority of *n*-type carriers, which the authors attributed to interfacial doping, while slight variations in density and mobility have been ascribed to variations in substrate quality and growth conditions.

Liu et al. removed the *n*-type doping by depositing F4-TCNQ *in situ* in UHV. The presence of the bulk bandgap was confirmed by an increase of about two orders of magnitude of the local resistance $R_L$ upon deposition of charge-depleting F4-TCNQ. Evidence for edge transport was provided by comparing resistances in local and non-local ($R_{NL}$) geometries using a Hall bar shaped device, as shown in Figure 3(a). In the non-local geometry, a very small resistive signal is expected, with $R_{NL}/R_L \approx 10^{-3}$. By increasing F4-TCNQ coverage, the Fermi level was shifted towards and into the bulk bandgap of Na$_3$Bi, with a corresponding increase of $R_{NL}$ which became comparable to $R_L$ in the absence of a magnetic field. In the edge state regime current flowing around the edges of the sample reaches the non-local voltage probes, and a large voltage is expected, with $R_{NL}/R_L \approx 1$, providing evidence of the existence of an edge state.[136] Resistive leakage into the bulk was observed as the temperature was increased above 20 K. From the magnitude of the resistance in the edge state regime, the authors estimated a mean free path of ~100 nm.

Figure 9(c) shows the magnetoresistance for the as-grown *n*-type doped bulk-conducting Na$_3$Bi (black line) and after F4-TCNQ deposition in the edge state regime (red line). The as-grown film shows a small positive magnetoresistance due to a mix of WL and WAL as expected for a gapped Dirac system. After F4-TCNQ deposition the film shows a giant negative magnetoresistance (GNMR) of about −80% at a magnetic field *B* = 0.9 T oriented perpendicular to the film.

This GNMR behaviour was modelled by developing a Boltzmann transport theory[137,138] for the spin-polarized helical edge states in the presence of solely magnetic impurity scattering. This is considerably more complex than previous treatments, in which the magnetic component of the scattering amplitude was treated perturbatively. Scattering off magnetic impurities with spin $\vec{S}$ was considered to be coupled to the edge spins via local exchange interaction; the effect of spin-conserving transitions (forward scattering) was averaged out, and single-particle spin-flip terms were determined in the Born approximation retaining linear terms in the impurity density. The resulting equation for GNMR is:

$$\frac{R(B) - R(0)}{R(0)} = \frac{\langle S_z \rangle}{S} \left[ \coth\frac{x}{2} - \frac{\frac{x}{2}}{\left(\sin\frac{x}{2}\right)^2} \right] - \tanh^2\left(\frac{1}{2}x\right)[2\,sech(x) + 1] \quad (5)$$

Here, $x = g\mu_B B / k_B T$, where *g* is the impurity gyromagnetic ratio, the sole material-dependent parameter, and $\langle S_z \rangle / S$ is given by the Brillouin function. In this description, Zeeman splitting of the impurity spin states due to the magnetic field causes flipping of the spin to become inelastic and therefore less favourable, leading to a negative value for the MR. In the case where spin-preserving scattering is allowed (i.e. a non-helical metal), the same formula applies but with a prefactor of $2/3$, leading to a maximum −67% magnetoresistance. Note that the perpendicular magnetic field does not open a gap in the edge state dispersion.

The blue dashed line in Figure 9(c) shows a fit to Eqn. (5), with fit parameter *g* = 15. The green dashed line shows the non-helical model (spin-preserving scattering allowed) with the same *g* = 15. The helical model fits the data well. Taking the prefactor in Eqn. (5) as a fit parameter provided an estimate of the amount of spin scattering in the edge state of >96%. The excellent fit of experiment to the model provides the first quantitative evidence of helical edge transport in a 2D topological insulator. Surprisingly, this effect has not been observed in other 2D topological insulators such as in HgTe quantum wells and WTe$_2$, because of the small topological bandgap in HgTe quantum wells, where Landau quantization effects become very important at low fields, driving the system towards a conventional insulator.[135] As for WTe$_2$, it lacks a horizontal mirror plane, and a perpendicular magnetic field opens a gap in the edge spectrum, leading to large *positive* magnetoresistance.[131,139]

# 5. Prospective applications

## 5.1. Devices

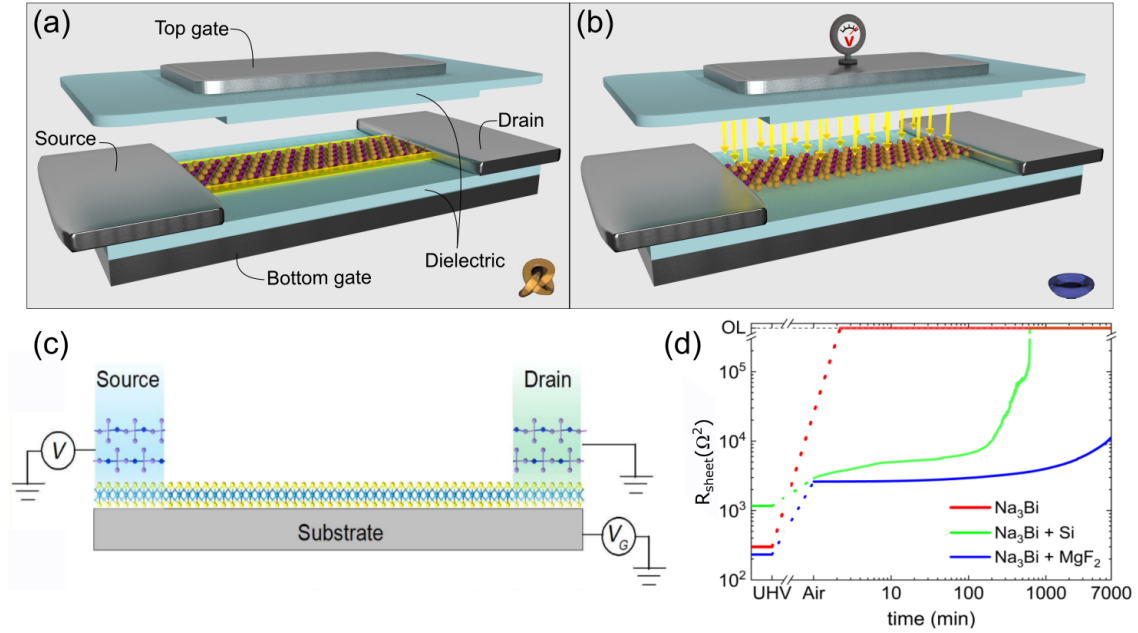

*Figure 10: Proposed design of a Na$_3$Bi-based topological transistor. (a) "ON" mode, with topological edge conduction between source and drain. (b) "OFF" mode, achieved by the application of an electric field to create a conventional insulator. (c) Proposed model of a TMDC-based transistor, using Na$_3$Bi as an electrical contact. (d) Resistivity of uncapped Na$_3$Bi (solid red line), Si- (solid green line) and MgF$_2$- capped (solid blue line) films on Al$_2$O$_3$ substrates as a function of time exposed to ambient atmosphere. Panel (c) adapted with permission from Cao, L. at al., Phys. Rev. Appl. **13,** 054030 (2020) . Copyright (2020) by the American Physical Society. Panel (d) adapted with permission from Liu, C. et al., ACS Appl. Mater. Interfaces, 12 (31), 35542–35546 (2020). Copyright (2020) American Chemical Society.*

Since their discovery, several proposals have been put forward to exploit the electronic properties of TI and TDS for the creation of topological devices.[140,141] Here, we propose some potential applications specifically tailored to Na$_3$Bi.

*5.1.1 Topological Transistors.* The creation of a TI-based field effect transistor (FET) is based on the concept that by breaking one of the relevant symmetries (e.g., sublattice symmetry, inversion symmetry, time-reversal symmetry or gauge symmetry) one can induce a transition from a topological insulator state, with ballistic edge transport, to a trivial (non-conducting) insulator. Electric-field induced switching between the two states has been proposed variously by exploiting the Rashba effect,[5] the Stark effect,[70,128,142–144] breaking inversion symmetry[145] or altering the sublattice potential.[5,146,147] Other proposals to control topology include tuning the tunnelling between the two topological edges,[148] exploiting backscattering from impurities to induce an "off" state,[149] or using time dependent electric fields,[150] magnetic fields,[132,151] or strain.[152,153] In particular, Yang and co-workers[152] proposed a piezo-topological transistor device by application of uniaxial strain to the Dirac semimetals Na$_3$Bi and Cd$_3$As$_2$: they demonstrated that a small applied strain can be used to control the quasiparticle spectrum in these materials, that a larger strain can trigger a topological phase transition, and that a non-homogenous strain profile can induce effects analogous to a warped space-time, with analogues in astrophysical objects like black/white holes.

Electric-field switching of the topological state has been successfully accomplished for Na$_3$Bi (see Section 4.2), paving the way for the realization of actual transistors based on this TDS. Ultrathin Na$_3$Bi has additional desirable properties: The bandgaps of topological and trivial states are larger than $k_B T$ at room temperature, the large topological bandgap is relatively insensitive to layer thickness, and Na$_3$Bi films are freestanding on technologically relevant materials such as silicon and even silicon oxide,[103] the two essential ingredients of current CMOS technology. We propose a possible design in Figure 10: the 2D topological layer is sandwiched between dielectrics in a double-gate configuration, where one of the two gates is required to generate a large electric field, enabling the switch between the "on" [panel (a)] and "off" states [panel (b)], and the other is used to control the carrier density so that the Fermi level is kept within the bulk bandgap. Previously, SiO$_2$ has been used as a bottom gate for Na$_3$Bi thin films, however it appears that the Na degrades the efficiency of the SiO$_2$ layer. Si appears not to interact with Na$_3$Bi, making intrinsic Si a potentially suitable dielectric medium. Other possibilities include hexagonal boron-nitride, SiN, or InP, which has been used to grow 3D topological insulators.[154] Creating a top dielectric and gate electrode is an additional challenge. MgF$_2$ is an excellent dielectric material that has demonstrated efficacy as a capping passivation layer. In this respect, significant progress has recently been made (see section 5.2.1).[58] Ionic liquids,[155,156] showing a high modulation performance in many other devices, offer another option for top gating, allowing for the Na$_3$Bi films to be safely transferred into a glovebox to protect against ambient exposure. While ionic liquids are unlikely to be useful in integrated circuits, they could serve to build more advanced proof-of-concept devices. Future work should focus towards the miniaturization of the proposed geometry, to realize a device on the scale of the electron mean free path in the material, thus minimizing the effect of defects and impurities.

A recent study by Cao et al.[157] explored a different approach to the design of a 2D electronic device, where the topological insulator Na$_3$Bi is integrated as an electrode contact for other 2D materials (namely, graphene, MoS$_2$, WS$_2$) as reported in Figure 10(c), as the Schottky-barrier height formed at the interface of Na$_3$Bi and the 2D semiconductors is reported to be lower than the one formed by bulk metals. In this case, the gate-tunability of the topological nature of Na$_3$Bi is expected to be exploited as a control for the design of these devices.

*5.1.2 Photodetectors.* Dirac semimetals have direct electronic transitions at all energies, which makes them uniquely applicable to long-wavelength (infrared and terahertz) photodetectors, while simultaneously exhibiting high mobility due to their high charge carrier velocities and absence of backscattering. In this respect, graphene has been proposed in several works as an efficient element of photodetection devices, either as a pure monolayer[158–160] or integrated in waveguide or microcavities to enhance its photoresponsivity.[161,162] In topological Dirac semimetals electrons disperse linearly as in graphene, with the further advantage of crystal symmetry protecting against an opening of the bandgap; moreover, their bulk nature potentially offers enhanced responsivity in terms of optical absorption compared to monolayer graphene. For this reason, the TDS Cd$_2$As$_3$ has already been proposed as the photosensitive element of ultrafast and broadband photodetecting devices,[163,164] capable of operating even at room temperature.[165,166] Based on the same principles, photodetecting devices have been realized with the topological insulator $Bi_2Te_3$[167] and the type-II Dirac semimetal $PtTe_2$.[168] Presently photodetection from a $Na_3Bi$ device remains unreported, but we anticipate that the linearity of the electron dispersion, high mobility of the carriers, and demonstrated growth of large-area crystals stabilised with optically transparent capping-layers, make $Na_3Bi$ a suitable platform for photodetection applications.

*5.1.3 Spin injectors*. Spintronic devices such as magnetic tunnel junctions rely on the injection of spin polarized current in the free layer from a contiguous spin injector, a material capable of converting charge current into spin current. Heavy spin-Hall metals like Pt and β-W have been proposed to this purpose, but they suffer from a low (<10%) charge to spin conversion.[169] The discovery and experimental realization of topological insulators, with their spin-momentum locked transport, has opened up new possibilities to realize efficient spin injectors. A charge to spin conversion as high as 36% is reported for $Bi_2Se_3$, but its low conductivity limits applications.[169] Conversely, a recent theoretical study suggests that the large spin-orbit coupling in $Na_3Bi$ can result in a conversion rate beyond 27%, while offering a resistivity 12.5 times lower than $Bi_2Se_3$.[170] For this reason, $Na_3Bi$ stands as an attractive candidate as the spin injector for the realization of efficient and fast magnetic tunnel junctions.

## 5.2. Challenges

*5.2.1 Air stability* The main challenge in the concrete and widespread utilization of $Na_3Bi$-based topological switches is the extreme air sensitivity of this material, whose surface oxidizes and deteriorates irreversibly within seconds of air exposure. Very recently Liu et al.,[58] following the capping approach developed for $Bi_2Te_3$ and $Bi_2Se_3$,[171–173] demonstrated the effectiveness of Si and $MgF_2$ layers as passivating agents: Si was chosen due to its demonstrated low degree of interaction with $Na_3Bi$ as a substrate, while $MgF_2$ is optically transparent and possesses a high dielectric constant. In their proposed approach, $Na_3Bi$ is grown via MBE in a Hall-bar configuration on a pre-patterned substrate, through a stencil mask. After the removal of the mask, Si or $MgF_2$ are deposited on the bar and the whole surface. For both capping layers the mobility as a function of carrier density is reported to be moderately lower, but not quenched – ruling out a chemical interaction between the passivating layer and the device. The roughness observed via SEM images of the $MgF_2$/$Na_3Bi$ surfaces suggests that the capping is imperfect, with micro-cracks or pinholes that might serve as preferential sites to initiate oxidation of the sample. Despite the fact that further work is required to address this issue, the $MgF_2$ capped samples exhibited an exceptional resistance to air exposure, as demonstrated in Figure 10(d), which reports the sheet resistance as a function of time after removal from the ultra-high vacuum growth chamber. The resistivity of the bare $Na_3Bi$ Hall bar reaches the GΩ range immediately, indicating significant degradation; the Si capped film after a slow initial deterioration shows a rapid increase of the resistance after about two hours, while the $MgF_2$ capped material shows a slow increase and a metallicity that persists in excess of 100 hours. This indicates that preservation of the unique electronic properties of Na$_3$Bi in both thin film form as a 3D TDS and ultra-thin form as a 2D TI is possible and can be improved in the future by improvement in the uniformity of the passivation layer.

*5.2.2 Scaling* Ultra-thin Na$_3$Bi as a topological insulator with large bulk band gap suggests a potential for ballistic transport on edge states at high temperatures (e.g. room temperature) which requires a smaller device channel size than the mean free path, which in turn requires micro- and nano- patterning. Conventional techniques employing photolithography and etching are presently incompatible with Na$_3$Bi due to the inevitable atmospheric exposure. One possibility for micro- or nano-scale device fabrication would be to use micro-fabrication techniques to pattern physical masks, similar to the large-scale Hall bar mask demonstrated in Ref. [154]. However, this scheme entails significant fabrication challenges. The device geometry could be significantly affected by the different positioning of the deposition sources during film growth, and control of the geometry of the film will require stringent control of the thickness, flatness, and relative thermal expansion of the mask and substrate. Aspects of these challenges are being addressed by the development of different stencil mask materials to achieve nanometer-scale patterning resolution for a wide variety of applications.[174] Patterned Si$_3$N$_4$ membranes can achieve 15 nm feature size,[174] while silicon-wafer based masks have achieved wafer-scale size for the patterning of 2D materials,[175] two promising avenues for future exploration.

Finally, to retain a clean substrate for film growth, present studies have used pre-deposited Ti/Au contacts which are also usually patterned using a metal shadow mask to avoid the residue of photoresist and other polymers. This results in edges of Ti/Au contacts which are not well shaped at the micro- or nano-scale, with rough, diffuse boundaries. In order to scale ultra-thin Na$_3$Bi to smaller dimensions, new techniques for fabricating contacts with precision at the micro- and nano-scale while retaining clean substrates suitable for MBE will be needed, however this should be feasible within the confines of conventional silicon processing, especially if semi-insulating silicon may be used as a substrate for further epitaxial growth. Other approaches could make use of the new technology for dry transfer of 2D materials and fabrication of van der Waals heterostructures, in which atomically clean insulators such as hexagonal boron nitride are routinely integrated with clean electrode material such as graphene.

## 5.3 Outlook

The challenges to fabricating mesoscopic ultrathin Na$_3$Bi topological devices are significant, but not insurmountable. At the moment there are a few alternatives for 3D TDS materials, but no demonstrated alternatives for 2D topological insulators with large bandgaps (significantly greater than room temperature) with demonstrated topological electronic transport properties. We expect that the drive to use topological insulators in devices will continue, and research into ultrathin Na$_3$Bi will advance with demonstrations of model devices coming in the near future, together with the investigation of more complex physics such as non-linear transport properties. If topological devices can be shown to be promising compared to their conventional counterparts, then these will provide strong motivation for the significant remaining challenges of developing robust scalable processes for growth and device fabrication.